\documentclass[reprint,aps,prd,amssymb, amsmath, nofootinbib]{revtex4-2}

\usepackage{graphicx}
\usepackage{subfigure}

\usepackage{mathtools}
\usepackage[svgnames]{xcolor} 

\usepackage[colorlinks=True, linkcolor=SlateBlue,            citecolor=SteelBlue,urlcolor=BlueViolet]{hyperref}
\usepackage[capitalise]{cleveref}
\usepackage{orcidlink}

\usepackage{aas_macros}

\newcommand{\beq}{\begin{equation}}
\newcommand{\eeq}{\end{equation}}
\newcommand{\unit}[1]{\ensuremath{\, \mathrm{#1}}}

\begin{document}

\title{Uncovering Stealth Bias in LISA observations of Double White Dwarf Binaries due to Tidal Coupling}
\author{Grace Fiacco\,\orcidlink{0009-0000-8844-982X}}
\author{Neil J. Cornish\,\orcidlink{0000-0002-7435-0869}}
\author{Hang Yu\,\orcidlink{0000-0002-6011-6190}}
\affiliation{eXtreme Gravity Institute, Department of Physics, Montana State University, Bozeman, MT 59717, USA}
\date{September 9, 2024}


\begin{abstract}\noindent
Double white dwarfs are important gravitational wave sources for LISA, as they are some of the most numerous compact systems in our universe. 
Here we consider finite-sized effects due to tidal interactions, as they are expected to have a measurable impact on these systems. Previous studies suggested that tidal effects would allow the individual masses to be measured, but there was a subtle error in those analyses. Using a fully Bayesian analysis we find that while tidal effects do not allow us to constrain the individual masses, they do yield informative lower bounds on the total mass of the system.
Including tidal effects is crucial to the accuracy of our estimation of the chirp and total mass. Neglecting tidal effects leads to significant biases towards higher chirp masses, and we see that the lower bound of the total masses is biased towards a higher value as well. For many systems observed by LISA, tidal effects can lead to a ``stealth'' bias, since only the first derivative of the frequency can be measured. To separate tidal effects from the usual point-particle decay we need to be able to measure the change in the second derivative of the frequency cause by the tides. This can only be done for high frequency systems observed with high signal-to-noise.
The bias, if not accounted for, can have significant astrophysical implications; for example, it could lead to an incorrect estimation of the population of potential Type Ia supernovae progenitors.
\end{abstract}

\maketitle


\section{Introduction}\label{sec:introduction}   

Since the first detection of gravitational waves by the Laser Interferometer Gravitational
Wave Observatory (LIGO)~\cite{Abbot_2016}, the field of gravitational-wave (GW) astronomy has flourished. Following LIGO's success, the GW community has been looking towards space-based detectors to expand the range of detectable frequencies and objects. In the coming decades, detectors such as the Laser Interferometer Space Antenna (LISA)~\cite{Amaro_2017} and possibly TianQin~\cite{Luo_2016} or Taiji~\cite{taiji} will be operating in the millihertz band, picking up new astrophysical signals such as merging massive black holes out to cosmological distances, and thousands of white dwarf binaries in our galaxy.

It is expected that for LISA, the vast majority of detectable signals will be from compact white dwarf (WD) binaries. These compact binaries emit almost monochromatic GWs with frequencies ranging from below 1 mHz up to around 20 mHz, which falls within LISA's sensitivity range of 0.1-100 mHz~\cite{Amaro_2017, Ruiter_2010}. LISA observations of compact galactic binaries can be used to constrain the evolutionary pathways of binary stars (for a review see Ref.~\cite{LISA:2022yao}), and their distribution throughout the galaxy~\cite{Adams:2012qw,Breivik:2019oar}. Most galactic binaries will be slowly evolving, with signals that can be well described by a frequency and frequency derivative $\dot{f}$~\cite{Shah_2012}. For sources with high enough frequencies, there is the potential to measure the second time derivative $\ddot{f}$ \cite{Nelemans_2004}, which can be used to probe finite-sized effects, and to infer physical properties about the binary. These effects depend on tidal interactions and mass transfer, which can potentially give us insight into the internal structure of white dwarfs such as their moments of inertia and component masses~\cite{Yu_2020}. 

\subsection{Double White Dwarfs}
\label{sec:dwds}
Double white dwarfs (DWD), or white dwarf binaries, are some of the most numerous detectable sources of gravitational waves in our universe. As only massive stars ($M\gtrsim10 M_{\odot}$) will collapse to a black hole or neutron star, the vast majority of stars will end their lives as a white dwarf. Nelemans~\cite{Nelemans_2001} has estimated that the Galaxy itself is populated with $\sim10^{8}$ DWD binaries. 

As these white dwarf binaries inspiral closer together due to GW emission, they can evolve into many types of variable systems such as AM CVn stars \cite{Nather_1981, Nelemans_2001}, rapidly rotating WDs~\cite{Warner_1998, VanZyl_2004}, and if paired with a neutron star, the WD can act as a donor to produce a millisecond pulsar~\cite{Bhattacharya_1991, Tauris_2011} to name a few. 

DWD systems are categorized into two types of binaries: detached and semi-detached DWDs. Detached DWDs are made up of simple binaries evolving closer together through gravitational wave emission. Semi-detached DWDs are more complicated, as mass-transfer occurs in these systems through Roche-lobe overflow from a hydrogen-deficient donor star to a more massive WD. These systems are observationally identified as AM CVn stars, producing EM counterparts from accretion in the form of X-rays, atomic line emission, and photometric variability \cite{Marsh_2011, Solheim_2010}. 

The vast majority of resolvable sources for LISA are expected to be from detached DWDs, with smaller numbers of AM CVn binaries, neutron star, and stellar mass black hole binaries~\cite{Nelemans_2001}. DWD systems are thought to be progenitors of Type Ia supernovae~\cite{Maoz_2014, Webbink_1984, Woosley_1986, Shen_2015, Iben_1984}, which LISA can help constrain, as their internal composition may be measurable through the tidal effects on the GW signature.  

These detached DWDs are also favored over semi-detached binaries as they tend to be more compact which creates a larger chirp mass, though both will be detectable by LISA. Detached DWDs will have a cleaner, stronger GW signal, whereas semi-detached systems will have Roche lobe overflow happening between the stars. When the lighter star overfills its Roche lobe, the masses of the system and the shape of the stars can both be varying in time, which in turn will vary the angular momentum of the system and potentially increase the orbit (leading to $\dot{f} < 0$) \cite{Nelemans_2004, Kremer_2017, Breivik_2018}. Assuming a stable mass transfer, the gravitational wave signal produced will be a simple anti-chirp, with a small negative frequency derivative due to their typically small chirp masses.

\subsection{Tidal Effects in DWDs}
\label{sec:tides}
Tidal interactions in detached DWDs can potentially lead to GW signatures and modifications that are detectable by LISA. These features are easiest to resolve in high frequency systems, starting at a gravitational wave frequency $f > 3$ mHz (or an orbital period of $P_{\rm orb} < 11$ min). The vast majority of DWDs emit at frequencies below 3 mHz, creating a confusion foreground. Assuming purely gravitational wave driven evolution on quasi-circular orbits, the number density of DWDs in our galaxy will scale as $dN/df \propto f^{-11/3}$, so at lower frequencies confusion noise cannot be neglected~\cite{Seto_2002,Timpano:2005gm}. Below 3 mHz only a small fraction of systems will be individually resolvable, while above 5 mHz essentially every DWD will be detectable and individually resolvable~\cite{Cornish_2013,Cornish_2017}. 

These high-frequency, detached DWDs are particularly interesting sources because they are measurably evolving in frequency. The higher the frequency the larger the frequency evolution, or ``chirp'', will be, which will allow us to also measure not just the first derivative, but the second derivative as well \cite{Nelemans_2004}, which can be used to distinguish between purely point-particle, GW driven evolution and tidal effects. This change in frequency comes partially from gravitational wave emission which shrinks the orbit and partially from the tidal coupling of the stars. 

Through gravitational wave emission, the binary loses angular momentum. This causes the orbital separation to decrease and the GW amplitude to increase, which leads to a measurable chirp signal \cite{BFSchutz_1996}. From this chirp, we can get fairly good measurements of the system's chirp mass $\mathcal{M}_{c}$, a combination of the component masses $m_{1}, m_{2}$ given as 
\beq
\mathcal{M}_{c} = \frac{(m_{1}m_{2})^{3/5}}{(m_{1}+m_{2})^{1/5}}. 
\eeq
The chirp mass dominates the frequency evolution of the signal during a binary's inspiral phase, which can be measurable for high-frequency systems over a long enough observation time \cite{Evans_1987}.

The change in frequency due to tidal coupling of the stars will give us constraints on the total mass through its dependence on the moment of inertia. To be able to properly measure the moment of inertia, we want systems that are either partially or totally tidally locked. As WDs evolve closer together, similar to binary stars, the tidal torques on the system will lead to tidal locking \cite{Zahn_1977}. The WD's spin $\Omega_{s}$ becomes well synchronized with the orbit $\Omega_{\rm orb}$, so $\dot{\Omega}_{\rm orb} \simeq \dot{\Omega}_{s}$. By assuming this synchronous rotation between the stars, it is possible to measure the frequency evolution well enough to extract information about the tidal effects \cite{Shah_2014}. This also places the stars in the traveling wave regime of internal oscillations, where very non-linear tidal effects occur \cite{Yu_2020,Fuller_2012,Fuller_2013,Fuller_2014}. Alternatively, studies have been done to measure the change in orbital period and braking index of these systems as a different way to infer information about tidal effects \cite{Piro_2011, Piro_2019}. Recently, Toubiana et al. 2024 \cite{Toubiana_2024} looked at similar effects, exploring in depth the impact of tidal torques and mass transfer on the detectability of these systems. 

We can only use these methods for determining the properties of DWDs and neutron stars. The presence of tides in these systems is what makes these compact binaries unique, as that is the key effect that will allow us to constrain their total mass. In contrast, black hole binaries have no surface and therefore no tides to synchronise their rotation.

\subsection{Previous Studies}
\label{sec:prevstudies}
Recently it was suggested~\cite{Kuns_2020,Yu_2020} that the moment of inertia of DWDs can be constrained to within 1\% using LISA observations. By relating the moment of inertia to the mass of each star, these constraints on the moment of inertia can be mapped to the constraints on the component masses. Wolz et al. 2021~\cite{Wolz_2020} expanded on this work by including tidal deformation effects to produce an empirical fit for the moment of inertia. They then carried out a Fisher Information Matrix analysis to determine how well the component masses of the system could be constrained~\cite{Wolz_2020}.

In this paper we expand upon these previous results, while also correcting errors in the previous analyses.
In Yu et al. \cite{Yu_2020}, the limits of integration in Eq. (72 - 74) were not adjusted to include tidal effects on the frequency evolution; they were computed considering only the point-particle effects. However, this should not affect the Fisher Matrix calculation to leading order. The error that contributed to the over-estimation in the detectability of $I_{\rm wd}$ was in the point-particle waveform onto which the tidal dephasing was added (Eq. 75). Here they chose a reference time in the stationary phase approximation \cite{Cutler_1994} that was many orders of magnitude larger than the observation time of 4 years, leading to a numerical instability. We expand upon the Wolz et al.~\cite{Wolz_2020} analysis, using their frequency evolution model combined with the frequency evolution term due to $I_{\text{wd}}$ from Yu et al.~\cite{Yu_2020}, and study these effects using a full Bayesian analysis.

We use a Markov Chain Monte Carlo (MCMC) method~\cite{Metropolis_1953, Hastings_1970} to perform a full Bayesian analysis on the binary system, applying techniques such as parallel tempering~\cite{Swendsen_1986} and differential evolution~\cite{636a7ce1ee074cbc8243cf136b4e3a08} to efficiently explore the posterior distribution for the model parameters. We find that, after accounting for the error in \cite{Yu_2020}, the component masses are no longer measurable for these systems from tidal effects. However, not including these effects in the analysis creates significant bias in the recovery of $\mathcal{M}_{c}$, shifting the value by as much as 10\%. Here, we use the term significant to describe when the value of $\mathcal{M}_{c}$ falls outside of the 99\% central credible interval of the posterior distribution.  

In Section \ref{sec:methods}, we describe the frequency evolution to the system, including finite-sized effects. We discuss different models for the moment of inertia and how those impact our analysis. We also discuss the concept of ``stealth'' biases and how that applies to this work (the term stealth bias was introduced in Ref.~\cite{Vallisneri:2013rc}). In Sections \ref{sec:results} and \ref{sec:discussion}, we present our results using a full Bayesian analysis with the above corrections included to the model. We discuss the differences between physical models and phenomenological models when including tidal effects, along with other effects that we decided to not include (i.e. tidal heating). In Section \ref{sec:conclusion}, we summarize and conclude.


\section{Frequency Evolution}\label{sec:methods}

The gravitational wave frequency (twice the orbital frequency for a circular binary) evolves due to a combination of gravitational and matter driven effects. Here we consider a combination of leading order post-Newtonian (PN) corrections and contributions due to tides through the moment of inertia $I_{\text{wd}}$ and tidal deformability $\Lambda$. We computed the modification to the moment of inertia due to the rotation of the star and found it to be negligible (see Appendix \ref{sec:appendix}), so we do not include it here. 

The following equations are written using geometric units ($G = c = 1)$.

\subsection{Post Newtonian and Tidal Effects}
\label{sec:fdot}

The dominant, 0PN point-particle frequency evolution for a quasi-circular inspiral obeys
\beq
\dot{f}_{\text{pp}} = \frac{96\pi^{8/3}}{5}\mathcal{M}_{c}^{5/3} f^{11/3},
\eeq
where $f$ is the gravitational wave frequency. Including the next order post-Newtonian correction to the point-particle dynamics yields~\cite{Blanchet_2020}
\beq
\dot{f} =\dot{f}_{\text{pp}} \left\{1 + \left(\frac{743}{336}-\frac{11\eta}{4}\right)(\pi M_{T}f)^{2/3}\right\}
\eeq
and 
\beq
\ddot{f} = \frac{11}{3}\frac{\dot{f}_{\text{pp}}^{2}}{f}\left\{1 + \frac{13}{11}\left(\frac{743}{336}-\frac{11\eta}{4}\right)(\pi M_{T}f)^{2/3}\right\},
\eeq 
where $M_{T}$ is the total mass. Here $\eta=m_1 m_2/ (m_1+m_2)$ is the symmetric mass ratio, which can be written in terms of chirp mass and total mass as $\eta = (\mathcal{M}_{c}/M_{T})^{5/3}$.
To add in the contributions from the tidal effects, we follow the prescription laid out by Yu et al.~\cite{Yu_2020} and Wolz et al.~\cite{Wolz_2020}. 

Including tidal effects, the frequency evolution takes the form
\beq
\dot{f} = \dot{f}_{\text{pp}}(1 + \Delta_{\text{1PN}}x + \Delta_{\text{I}}x^{2} + \Delta_{\Lambda}x^{5}).
\eeq
Here,
\beq
x = (\pi f M_{T})^{2/3},
\eeq
\beq
\Delta_{\text{1PN}} = (\frac{743}{336}-\frac{11\eta}{4}),
\eeq
\beq
\Delta_{\Lambda} = (39/8)\Tilde{\Lambda},
\eeq
\begin{multline}
 \Tilde{\Lambda} = \frac{8}{13}[(1+7\eta - 31\eta^{2})(\Lambda_{1} + \Lambda_{2}) \\+ \sqrt{1-4\eta}(1+9\eta-11\eta^{2})(\Lambda_{1} - \Lambda_{2})], 
\end{multline}
and 
\beq
\Delta_I = \frac{3 \sum_i^{1, 2} I_{{\rm wd}, i}/(\eta M_{T}^3) }{1- 3 \sum I_{{\rm wd}, i} x^2 / (\eta M_{T}^3) },
\eeq
where we explicitly show the sum over the component moments of inertia. We have factored out the $x^2$ in the numerator to stay consistent with the format of (5). The above expressions are all found in Wolz et al. \cite{Wolz_2020} except for $\Delta_{I}$, which is found in Yu et al. \cite{Yu_2020}.

We consider two models for the dimensionless tidal deformability, $\Lambda_{1(2)}$ depending on our model for $I_{\text{wd}}$ (see Section \ref{sec:Iwd}). When using the model  given by Kuns et al. 2020~\cite{Kuns_2020}, we use the simpler equation
\beq
\Lambda_{1(2)} = \lambda_{1(2)}/m_{1(2)}^{5}.
\eeq
$\lambda$ relates the star's induced quadrupole moment $Q_{ij}$ to the external tidal field. $\lambda$ is also related to the Love number $k_{2}$ by $k_{2} =(3/2)\lambda/R^{5}$ \cite{Hinderer_2008}. $R$ is the radius of the WD (in units of $m$), and $k_2\simeq 0.01-0.1$ for polytropes with an adiabatic index of $\Gamma=4/3-5/3$, where $\Gamma$ determines the equation of state as $P\propto \rho^\Gamma$. Here, $P$ and $\rho$ represent pressure and density respectively.

To determine the radius for each WD, we use Equation 3 from \cite{Kuns_2020}, with $R(m)$ converted to units of $Gm/c^{3}$:
\beq\label{kunsradius}
R(m) = 0.033\left(\frac{m}{0.6M_{\odot}}\right)^{-1/3} \unit{s}.
\eeq

When using the fit given by Wolz et al. \cite{Wolz_2020}, we use their corresponding fit for $\Lambda$:
\beq
\text{ln}\Lambda = 2.02942 + 2.48377 \text{ln}\bar{I}.
\eeq
$\text{ln}\bar{I}$ is the log of the dimensionless moment of inertia, and is defined in Equation 16 (see Section \ref{sec:Iwd}).

Taking another time derivative, we find
\begin{multline}
\ddot{f} = \frac{11}{3}\frac{\dot{f}_{\text{pp}}^{2}}{f}(1 + \frac{24}{11}\Delta_{\text{1PN}}x + \frac{26}{11}\Delta_{\text{I}}x^{2}\\ + \frac{19}{11}\Delta_{\text{I}}^{2}x^{4} + \frac{32}{11}\Delta_{\Lambda}x^{5} + \frac{21}{11}\Delta_{\Lambda}^{2}x^{10}).    
\end{multline}
Here we leave out the $\Delta_{\text{1PN}}^{2}$ term due to it not being measurable, but keep the higher order $\Delta_{\text{I}}, \Delta_{\Lambda}$ terms in our $\ddot{f}$ expression. These higher order terms appear due to the substitution of $\dot{f}$ in to our $\ddot{f}$ expression when we take the time derivative.

\begin{figure}[h!]
    \centering
    {\includegraphics[width=0.45\textwidth]{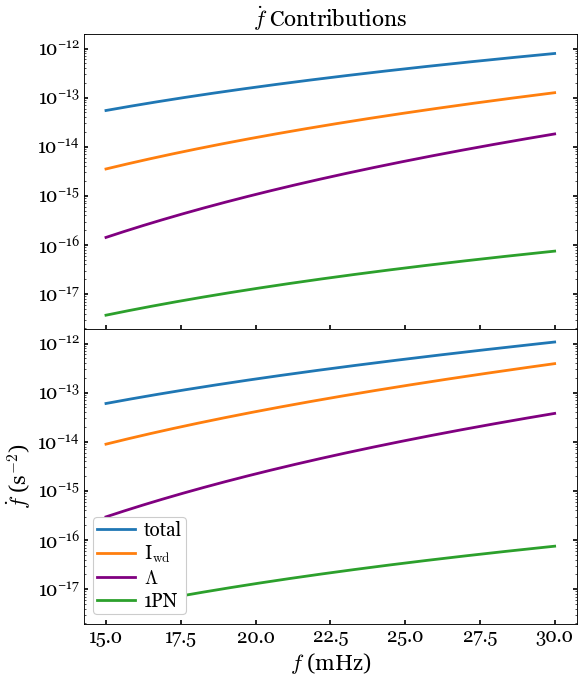}}
    \caption{We show the contributions to $\dot{f}$ from the 1PN correction, the moment of inertia, and the tidal deformability. The top panel is $\dot{f}$ using the Wolz et al.~\cite{Wolz_2020} model (Equations (12) and (16)); the bottom panel uses the model given by Kuns et al.~\cite{Kuns_2020} (Equations (10) and (15)). We see that the contribution from the moment of inertia is the most significant out of the three. We expect to see this term affecting our resulting posterior distributions the most, whereas the tidal deformability and 1PN correction will not change them as significantly.}
    \label{fig:fdot plot}
\end{figure}

To further see the importance of each contribution to $\dot{f}$, we plot the different tidal contributions and the 1PN contribution in comparison to the total $\dot{f}$ in Figure \ref{fig:fdot plot}. From here, we can see that though we include both the $\Delta_\Lambda$ and $\Delta_{1\text{PN}}$ terms, they do not contribute significantly to our overall $\dot{f}$. We do see that the expression for $\Delta_\Lambda$ further decreases the value for the total $\dot{f}, \ddot{f}$ when using the empirical relation from Wolz et al.~\cite{Wolz_2020}. The main contributor here is $\Delta_{\text{I}}$, so we expect to see this term affecting our posterior distributions more than the other two terms will.

\subsection{Moment of Inertia $I_{\text{wd}}$}
\label{sec:Iwd}
We looked at two different models for the moment of inertia of a WD in this paper; a simple power-law fit derived by Kuns et al. \cite{Kuns_2020} and an empirical fit from Wolz et al. \cite{Wolz_2020}. The power-law model treats the WD as a non-relativistic polytrope with $P\propto \rho^{5/3}$ \cite{Chandrasekhar_1931, Balberg_2000}, whereas the empirical fit describes a more realistic WD at zero temperature, which is a good approximation for more massive WDs \cite{vanKerkwijk_2005}.

The power-law model takes the form
\beq\label{KunsIwd}
I(m) = 3.1\times10^{50}\left(\frac{m}{0.6M_{\odot}}\right)^{1/3} \unit{g}\unit{cm^{2}}
\eeq 
whereas the empirical fit has the form 
\begin{multline}
    \text{ln}\Bar{I} = 24.7995-39.0476\hat{m}_{1(2)} + 95.9545\hat{m}_{1(2)}^{2} \\- 138.625\hat{m}_{1(2)}^{3} + 98.8597\hat{m}_{1(2)}^{4} - 27.4000\hat{m}_{1(2)}^{5}
\end{multline}
where $\hat{m}_{1,2}\equiv m_{1,2}/M_\odot$  are the unitless WD masses.

\begin{figure}[h!]
    \centering
    {\includegraphics[width=0.45\textwidth]{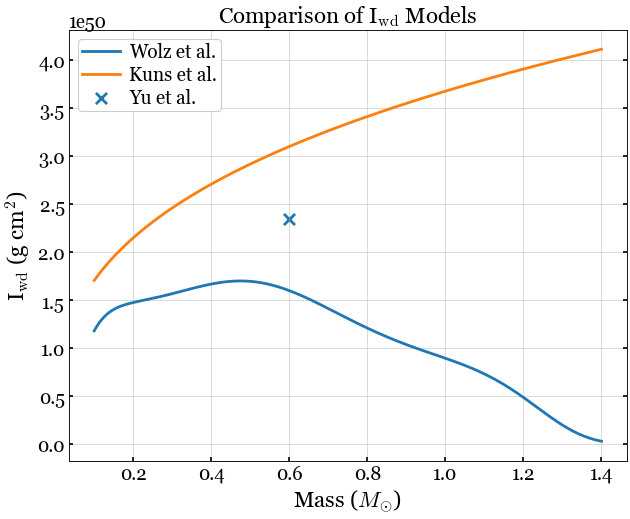}}
    \caption{Moments of inertia as a function of mass. The top curve is the power-law fit from Kuns et al. \cite{Kuns_2020}, while the bottom curve is the empirical fit given by Wolz et al.~\cite{Wolz_2020}. At high masses, the power-law fit over-estimates $I_{\text{wd}}$, which inflates the detectability predictions. The empirical fit underestimates $I_{\text{wd}}$ at low masses, but for high masses it is the more accurate model, using a zero-temperature approximation. We also include the value of $I_{\text{wd}}$ used in \cite{Yu_2020} for further comparison, generated from the $\tt{MESA}$ stellar evolution code.}
    \label{fig:Iwd comp plot}
\end{figure}

Figure \ref{fig:Iwd comp plot} shows the comparison between the two models. There are significant differences between the power-law fit and the empirical fit; the two models differ by a factor of 3 at the highest mass and we see that the empirical fit does not monotonically increase, which creates an interesting multi-valued effect in our analysis. We also include the moment of inertia used by \cite{Yu_2020} for completeness, as they do not use any of the relations specified above. They used the stellar evolution code $\tt{MESA}$~\cite{Paxton_2021} to generate a WD model, which gives a moment of inertia value between the two models we consider for a WD of mass $0.6M_{\odot}$.

From here we can see that the power-law fit starts over-estimating the moment of inertia for WDs above 0.6$M_{\odot}$, which creates an artificial boost in detectability. The method used by \cite{Yu_2020} over-estimates $I_{\text{wd}}$ as well. Though at masses $<0.6M_{\odot}$ the empirical fit underestimates $I_{\text{wd}}$, at higher masses the fit is more accurate. That region is where the zero-temperature approximation becomes the better approximation to use, and this is the mass range we need to detect tidal effects in WDs. 

\subsection{Parameter Restrictions}
\label{sec:param_restrictions}
When starting our analysis, we need to be aware of the astrophysical restrictions for each of our parameters. Since our system is a detached DWD, effects such as mass transfer need to be avoided when setting up the analysis. We must also consider properties such as the maximum and minimum masses for the individual stars, and the maximum frequencies that we can explore before these stars either begin to mass transfer or collide and merge.

\begin{figure}[h!]
    \centering
    {\includegraphics[width=0.45\textwidth]{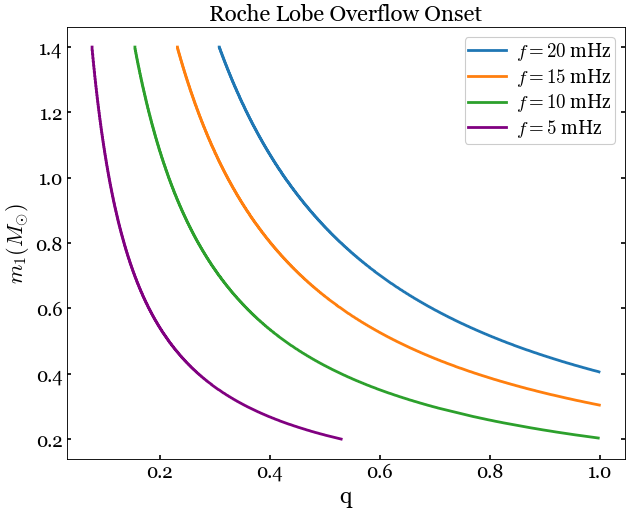}}
    \caption{Curves in ($m_{1}, q)$-space where Roche lobe overflow sets in for a range of GW frequencies. Points to the lower left of each curve represent $(m_{1}, q)$ pairs that have overflow happening. We can reach more unequal mass ratios the lower in frequency we go, as the stars are farther apart than at higher frequencies.}
    \label{fig:roche plot}
\end{figure}

We first set upper boundaries for $(m_{1}, m_{2})$ so they do not exceed the Chandrasekhar limit \cite{Chandrasekhar_1994}. We then need to restrict the allowed lower masses to avoid Roche lobe overflow. To find where this begins, we use Eggleton's approximation \cite{Eggleton_1983} to find the Roche lobe radius, $R_{\text{L}}$:
\beq\label{rochelobe}
R_{\text{L}} = a_{\rm orb}\frac{0.49q^{2/3}}{0.6q^{2/3} + \text{ln}(1+q^{1/3})},
\eeq
where $q = m_{2}/m_{1}$ is the mass ratio and $a_{\rm orb}$ is the binary separation, which is a function of the total mass of the system and the orbital frequency. To determine when this boundary is reached, we compare $R_{\text{L}}$ to the radius of the lighter star, $R_{2}$. This is calculated using the relation provided in Equation 12. Once the lighter star's radius exceeds $R_{\text{L}}$, mass transferring begins, so we restrict our analysis to systems where $R_{2} < R_{\text{L}}$. Figure \ref{fig:roche plot} shows at which mass combinations this overflow happens for a range of GW frequencies. The region to the lower left of each curve represents the $(m_{1}, q)$ pairs where Roche lobe overflow happens. We can see that for low frequencies, we can reach highly unequal mass ratios before overflow begins, due to the wider orbital separation between the stars. The maximum mass ratio that can be reached gets smaller as we increase in frequency and the orbital separation shrinks.

This approximation not only constrains the allowed masses for the lighter companion, but it also governs the maximum frequency we that consider. For our system of a $(0.6-0.8)M_{\odot}$ binary at a GW frequency $f = 20$ mHz (corresponding to an orbital frequency of $f_{\text{orb}} = 10$ mHz), the system has not yet exceeded its Roche lobe. If we increase the frequency past $f = 25$ mHz, the system enters the mass transferring regime. This also depends strongly on the mass ratio $q$; as $q$ decreases, corresponding to a highly unequal-mass binary, the lower the frequency of the system must be to avoid mass transfer.

\subsection{Frequency Parameterizations}
\label{sec:freq_parameterization}
To perform parameter estimation on the frequency evolution models outlined above, we use a full Bayesian analysis code instead of a simpler Fisher Information Matrix analysis, though calculating the Fisher Matrix is still helpful as it provides us with analytical insights. To make the calculations more numerically stable, as $\dot{f}$ and $\ddot{f}$ are quantities on the order of $10^{-13} \unit{s^{-2}}$ and $10^{-24} \unit{s^{-3}}$, we parameterize the frequencies in terms of unitless quantities $\alpha = \text{T}_{\text{obs}}f, \beta= \text{T}_{\text{obs}}^{2}\dot{f}, \gamma= \text{T}_{\text{obs}}^{3}\ddot{f}$. Here, $\text{T}_{\text{obs}}$ is the observation time, which we take to be 4 years in this paper.  

To be able to distinguish between point-particle and finite-sized effects, we need to be able to measure not only $\ddot{f}$, but the difference between $\ddot{f}$ and the point-particle prediction, which can be expressed in terms of the measured $f$ and $\dot f$. In terms of the dimensionless variables $\alpha, \beta,\gamma$ the difference is given by
\beq
\delta\gamma = \gamma - \frac{11}{3}\frac{\beta^{2}}{\alpha} = \gamma_{\text{tide}} - \gamma_{\text{pp}}.
\eeq

Now we can look at the fractional errors $\Delta\delta\gamma / \delta\gamma$ to see if this contribution from the second derivative is measurable. To estimate this we compute the Fisher Information Matrix using a simple sinusoidal signal~\cite{Cornish_2003}
\beq
h(t) = A\cos{(\phi(t))}
\eeq
where $A$ is the amplitude, and $\phi(t)$ is the phase
\beq
\phi(t) = \phi_{0} +  2\pi \alpha \left( \frac{t}{T_{\text{obs}}}\right) + \pi\beta\left(\frac{t}{T_{\text{obs}}}\right)^{2} + \frac{\pi}{3}\gamma\left(\frac{t}{T_{\text{obs}}}\right)^{3}.
\eeq

The Fisher Matrix is defined as \cite{Seto_2002}
\beq
\Gamma_{ij} = \frac{2}{S_{n}(f)}\int^{T_{\text{obs}}}_{0} dt \frac{\partial h}{\partial \gamma_{i}}\frac{\partial h}{\partial \gamma_{j}}
\eeq
where $S_{n}(f)$ is the noise spectral density and $\gamma_{i}$ are the parameters $(A, \phi_{0}, \alpha, \beta, \gamma, \delta\gamma)$. We get the estimated errors on our parameters by inverting the full Fisher Matrix \cite{Cutler_1994}, 
\beq
\sqrt{\langle(\Delta\gamma_{i})^{2}\rangle} = \sqrt{\Sigma_{ii}},
\eeq
with $\Sigma \equiv \Gamma^{-1}$. It is important to use the full inverse when calculating the errors on our parameters here, as the parameters are highly correlated. By using the full inverse, we ensure that our error estimations include information from the cross-correlations between the different parameters. 

\begin{figure}[h!]
    \centering
    {\includegraphics[width=0.45\textwidth]{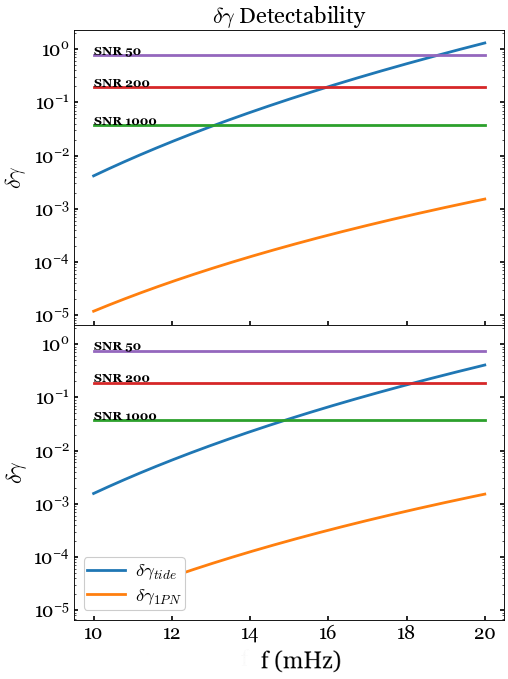}}
    \caption{We compare $\delta\gamma$ for 1PN corrections and tidal corrections using both models of $I_{\text{wd}}$ for a range of SNRs. The top plot uses the power-law model. The bottom plot uses the empirical model. We overplot the error $\Delta\delta\gamma$ from the Fisher Matrix analysis to determine when $\delta\gamma$ becomes measurable (i.e. when $\Delta\delta\gamma / \delta\gamma = 1$). At 1000 SNR, the tidal correction becomes measurable by 14 mHz using the power-law model, compared to 16 mHz using the empirical model. We see the frequency where $\delta\gamma$ becomes measurable increases as we decrease the SNR. By an SNR of 50, $\delta\gamma$ is no longer measurable using the empirical model and is only measurable at high frequencies using the power-law model. The 1PN correction is never measurable.}
    \label{fig:delta plot}
\end{figure}

Figure \ref{fig:delta plot} shows a comparison in detectability with and without including tidal effects. To properly compare the two models and their fractional errors, we plot the relationship between $\delta\gamma$ and the frequency, and overlay $\Delta\delta\gamma$, the parameter estimation uncertainty, to see where the fractional error reaches unity, determining the beginning of detectability.

Note that $\Delta \delta \gamma$ (and $\Delta \beta$) is approximately constant for a given SNR. The parameter $\gamma$ (and $\beta$) enters linearly into the waveform $h(t)$, so $\partial h/\partial \gamma$ (and $\partial h/\partial \beta$) becomes only a function of the SNR.  

From Figure \ref{fig:delta plot}, we can see that for an SNR of 1000, the tidal contribution becomes measurable at a GW frequency of 14 mHz using the scaling relation for $I_{\text{wd}}$ given by Kuns et al.~\cite{Kuns_2020}. When applying the Wolz et al. model~\cite{Wolz_2020}, the detectability further drops, with $\Delta\delta\gamma/\delta\gamma$ not reaching unity until past 16 mHz. 
For systems with lower signal-to-noise, departures from point-particle evolution can not be detected until higher frequencies, becoming unmeasurable below an SNR of 50.

\subsection{Stealth Biases}
\label{sec:bias}

When we have sources in LISA that are high enough in both frequency and SNR to be able to measure $\ddot{f}$, we can detect effects such as tides and other deviations away from the point-particle model. This will not be the case for the majority of LISA sources though, as the number of DWDs will increase as the frequency decreases \cite{Seto_2002}. 

The search for galactic binaries in simulated LISA data so far has been done using a phenomenological model (search parameters including $f, \dot{f}, \ddot{f}, A$ among others) \cite{Crowder_2007, Littenberg_2011}. The benefit for using this parameterization is its flexibility; it can accurately handle many types of models (tides, heating, mass transfer, etc.), but we then need to map this to a physical model to extract the physics. It is in this stage of the analysis that we can see how our model deviates from the point-particle model to detect any biases or features. 

If we have an incorrect model, detecting any deviations becomes much harder to do. These cases of incorrect modeling create biases in our analysis, and often exist due to the model only including point-particle contributions and lacking other contributions that are necessary to properly model the system. In our case, biases become apparent when we neglect the tides, leaving them out entirely. For low-frequency and SNR sources in which the tides are not detectable through $\delta \gamma$ in the first place, this is of particular concern, as that is where the majority of DWDs exist. 

In reality, detecting these biases from the data directly is very difficult; when we have systems where only the first frequency derivative is measurable, we have no way of knowing if our chosen model is correct, as we have no way to cross-check the data. We call these types of biases ``stealth'' biases, as they sneak into our data without our knowing and are governed by the first derivative instead of the second derivative~\cite{Vallisneri:2013rc}. 

\begin{figure}[h!]
    \centering
    {\includegraphics[width=0.40\textwidth]{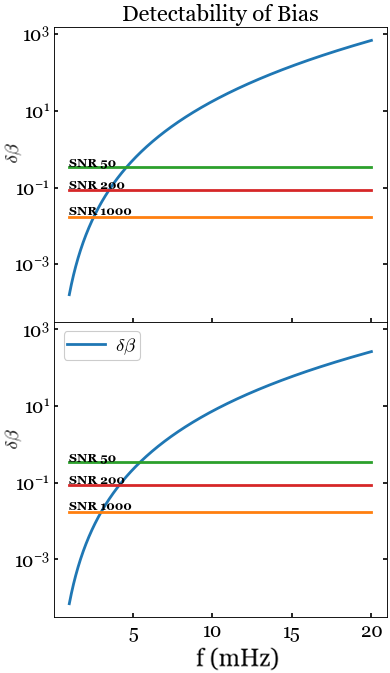}}
    \caption{We look where the Fisher Matrix errors $\Delta\beta$ cross the $\delta\beta$ curve at different SNRs. Above these crossing points we expect to see biases in our analysis. We see that for all SNRs ranging from below 50 up past 1000, we will start seeing a bias at $ f < $10 mHz, which is where the majority of LISA sources reside.}
    \label{fig:beta plot}
\end{figure}

Whereas previously we examined the detectability of $\delta\gamma$ to determine when we might see the effects of the tides in our data, here we turn to the detectability of $\beta$, as that is what governs our recovery of $\mathcal{M}_{c}$, which is where this bias will be most apparent. To explore how this bias could affect low-frequency and low-SNR sources, we look at the systematic error in $\beta$ compared to the statistical error given by the Fisher Matrix, $\Delta\beta$ (similar to Figure \ref{fig:delta plot} where we look at the systematic vs statistical error of $\delta\gamma$ to determine when tides are detectable). We define our systematic error in $\beta$ as
\beq
\delta\beta = |\beta_{\text{tide}} - \beta_{\text{pp}}|
\eeq
and when $\delta\beta \sim 3\Delta\beta$, or outside the 3-sigma range, we should start seeing a significant bias in our data. Figure \ref{fig:beta plot} shows this threshold. This can easily be scaled to other SNRs not listed, as $\Delta\beta$ scales inversely as a function of SNR as
\beq
\Delta\beta = 17.08\left(\frac{1}{\text{SNR}}\right)
\eeq
where 17.08 is $\Delta\beta$ at an SNR of 1.

\begin{figure}[h!]
    \centering
    {\includegraphics[width=0.45\textwidth]{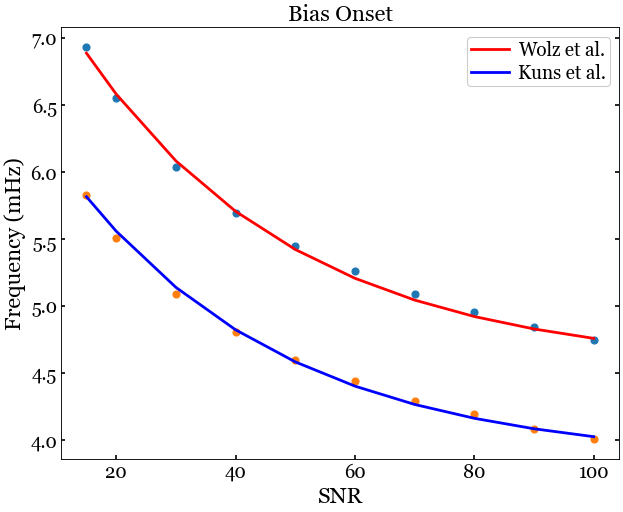}}
    \caption{For both $I_{\text{wd}}$ models, we fit the crossing frequencies as a function of SNR for when the stealth bias begins to show up (i.e. when the analysis is biased outside the 1-sigma range). For low frequency systems the bias persists even for high SNR sources.}
    \label{fig:bias fit plot}
\end{figure}

To further quantify this bias effect across different SNRs, we created a fit of the crossing frequency where the bias begins to show as a function of SNR, shown in Figure \ref{fig:bias fit plot}. This fit has the form
\beq
f_{\text{bias}}(x) = ae^{bx} + c
\eeq
where $x$ is the SNR. The fitting parameters for the empirical model are: a = 3.5738, b = -0.0275, c = 4.54. For the power-law model, the scaling parameters are: a = 3.0074, b = -0.0279, c = 3.84.

Once we can detect $\delta\gamma$ we can break this bias, as the tides are no longer negligible and will be included in the model. But as shown in Figure \ref{fig:delta plot}, this requires very high frequencies and SNRs. When we compare this to Figure \ref{fig:beta plot}, we see that this stealth bias first will show up at very \textit{low} frequencies and SNRs, where LISA sources will be numerous. 

This shows that there is a wide range of frequencies and SNRs where there will not be detectable tides but there will be a bias for neglecting them. Since very few LISA sources will be at a high enough SNR and frequency to detect $\delta\gamma$, this bias is extremely stealthy.


\section{Results}\label{sec:results}
We present several subsets of results in this paper. In Section \ref{sec:physparams}, we show our main results from using a physical parameterization. We sample directly in $(\mathcal{M}_{c}, M_{T})$ and post-process to obtain the rest of our parameters $(m_{1}, m_{2}, q)$. We see that the component masses are not measurable through tidal effects, but leaving these effects out creates large biases in the chirp and total mass. In Section \ref{sec:phenomparam}, we show the conclusions we have drawn from using the phenomenological parameterization described in Section \ref{sec:freq_parameterization}; resampling from frequencies to physical parameters is not accurate enough, as it misses data in key areas of parameter space. Section \ref{sec:bias2} discusses our key result about stealth biases, and shows just how significantly this affects our analysis and should not be ignored. 

All of the following results are for a white dwarf binary with component masses of $(0.8 - 0.6)M_{\odot}$, a frequency of $f = 20\unit{mHz}$, and at an SNR of 1000 unless otherwise stated, as we require high-frequency and high-SNR systems to be able to detect these tidal effects.  

\subsection{Constraining Physical Parameters}
\label{sec:physparams}
Here we use a physical parameterization for the waveform, $(\mathcal{M}_{c}, M_{T}, m_{1}, m_{2}, q)$. Since we are sampling in parameters $(\mathcal{M}_{c}, M_{T})$, but our prior is described in terms of parameters $(m_{1}, q)$ (or $m_{2})$, we need to apply a Jacobian transformation to account for this re-parameterization. This is done by computing 
\beq\label{jac}
J(\mathcal{M}_{c}, M_{T}) = \left|\left|\frac{\text{d}(m_{1}, q)}{\text{d}(\mathcal{M}_{c}, M_{T})}\right|\right|
\eeq
where we are taking the absolute value of the determinant of our Jacobian matrix. Since the prior is taken to be uniform in $(m_{1}, q)$ ,  the prior on the chirp mass and total mass is simply proportional to the Jacobian (\ref{jac}).

\begin{figure}[h!]
    \centering
    \begin{subfigure}[]
        \centering
        {\includegraphics[width=0.47\textwidth]{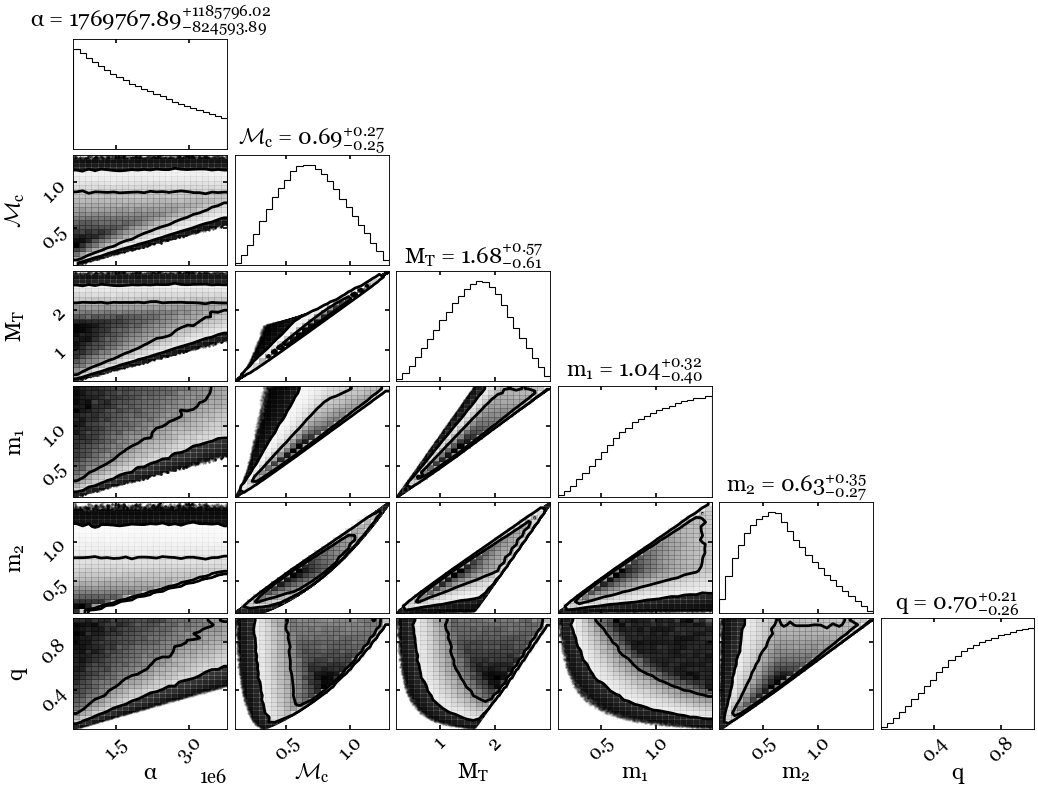}}
    \end{subfigure}
    \begin{subfigure}[]
        \centering
        {\includegraphics[width=0.47\textwidth]{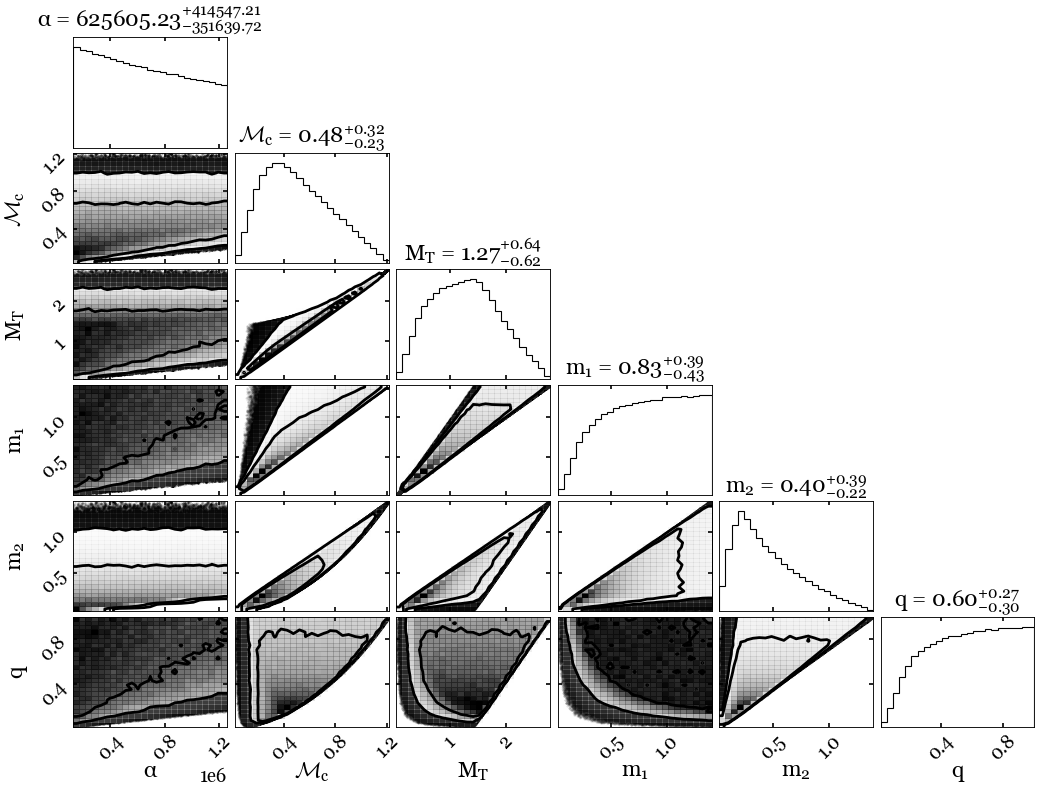}}
    \end{subfigure}
    \caption{The prior distribution for the parameterized frequency $\alpha$ and the mass parameters, initially uniform in $(m_{1}, q)$ at $f=20$ mHz (a) and $f=5$ mHz (b), where all masses are in units of solar mass ($M_{\odot}$). Implementing physical restrictions on the parameters due to the filling of the system's Roche lobe modifies the prior distributions significantly, but we see at low frequencies we recover uniform priors on $(m_{1}, q)$ at moderate masses and mass ratios. }
    \label{fig:prior plot}
\end{figure}

When including the constraints on our parameters (see Section \ref{sec:param_restrictions}), our priors, though starting as uniform, become non-trivial. Figure \ref{fig:prior plot} shows our prior distribution for initially uniform priors in ($m_{1}, q)$ at $f=20$ mHz and $f=5$ mHz. Implementing the lower limits on ($m_{1}, q)$ due to Roche lobe overflow significantly modifies our once uniform priors at $f=20$ mHz. Using Equations \ref{kunsradius} and \ref{rochelobe} at the onset of mass transfer, we can derive the expected relation between ($m_{1}, q$) in the low-$q$ limit to be $q \propto m_{1}^{-1}$. This is seen in the corresponding 2D histogram in Figure \ref{fig:prior plot}. We also see at $f=5$ mHz the priors are modified less from Roche lobe overflow, recovering the original uniform priors at moderate ($m_{1}, q)$ pairs. As the orbital separation of the binary is farther apart at lower frequencies, the binary must have a highly unequal mass ratio to begin mass transferring.

We first explored the effect of rotation of the binary and how it modified the moment of inertia. As the WDs become tidally locked and inspiral inwards, the rotation of each star must increase to maintain the tidal locking. This would cause the stars to deform, which in turn would modify the moments of inertia. We found that this deformation is negligible at the higher masses and spins we are considering in this paper. More information about our calculations can be found in Appendix \ref{sec:appendix}. 

As mentioned previously, we found two errors in the results presented by Yu et al. \cite{Yu_2020}. Their analysis was done in the frequency domain, where they calculate the phase $\Psi(f)$ of the GW waveform by integrating the frequency evolution from the point-particle and tidal contributions. One thing that they did not account for in their calculation was the adjustment of their limits of integration to also include the tidal effects, opposed to just the point-particle contribution. This, however, should not significantly effect their Fisher Matrix calculation to leading order. What was wrong in \cite{Yu_2020} was actually the point-particle waveform onto which the tidal dephasing was added. The point-particle part (Eq. (75) of \cite{Yu_2020}) was computed under the stationary phase approximation as in \cite{Cutler_1994}. While the expressions are in principal correct (the requirement on the duration of observation is $T_{\rm obs}^2 \dot{f}\gg 1$, which is satisfied for binaries at $f=20$ mHz and $T_{\rm obs}=4\,{\rm yr}$), this can lead to numerical instabilities if $t_c$ is poorly chosen. 

A significant portion of the point-particle phase shift can be removed with a more appropriate choice of $t_c$  (e.g., $t_c=-3/11 f/\dot{f}$), or by using the Taylor expanded version of the phase evolution as in Eq. 11 of \cite{Wolz_2020}, where $t_0=0$ is a natural choice. Ref. \cite{Yu_2020} did include $t_c$ as part of the Fisher Matrix analysis, which should achieve the desired linear detrending of $\Psi(f)$. However, the numerical accuracy required when $t_c$ is off by $\sim 4,000\,{\rm yr}$ is far beyond achievable. This caused the chirp mass $\mathcal{M}_c$ to be incorrectly well-constrained in \cite{Yu_2020}, which further underestimated the measurement uncertainty on the total mass $M_T$ as the two are highly correlated. With these fixes implemented, their result longer stands; Figure \ref{fig:delta plot} shows that at 10 mHz, this effect is never measurable.

The moment of inertia model itself is another aspect we explore. We consider two different moments of inertia models, described in Section \ref{sec:Iwd}. For high-mass WDs, the zero-temperature approximation is more accurate to use than for lower mass WDs $(< 0.6M_{\odot})$ \cite{Taylor:20}, making the empirical model for $I_{\text{wd}}$ the more accurate of the two we looked at. 

The choice of model effects the measurability of the tides as well. As shown in Figure \ref{fig:Iwd comp plot}, the power-law model over-estimates $I_{\text{wd}}$ by a factor of 3 at the high mass range. This over-estimation, combined with the over-estimations made by Yu et al.~\cite{Yu_2020} further inflates the detectability of the component masses.

Figure \ref{fig:delta plot} shows that the tidal effects become measurable around $f\simeq 14\unit{mHz}$ when using the power-law model, but it is not until $f\simeq 16\unit{mHz}$ that this is measurable when using the more accurate empirical fit. By comparison, we see that when only including the 1PN corrections to the frequency, $\delta\gamma$ will never be measurable for these systems. 

\begin{figure*}
    \centering
    \begin{subfigure}[]
        \centering
        {\includegraphics[width=0.75\linewidth]{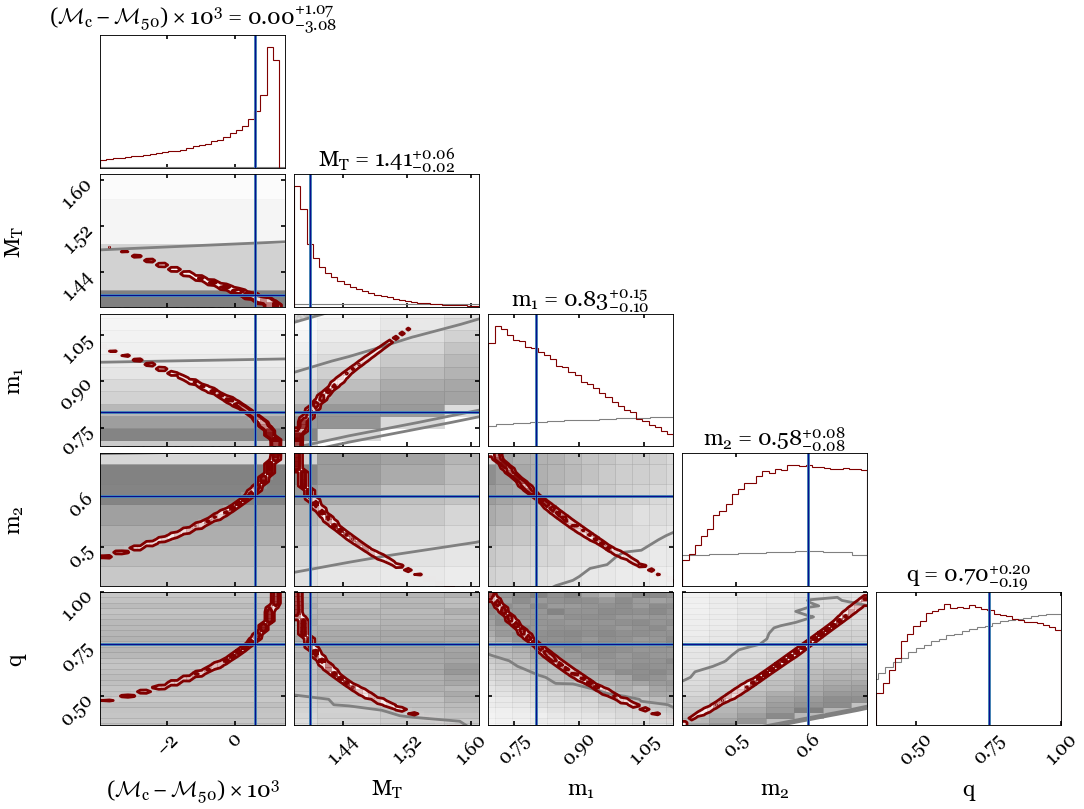}}
    \end{subfigure}
    \begin{subfigure}[]
        \centering
        {\includegraphics[width=0.75\linewidth]{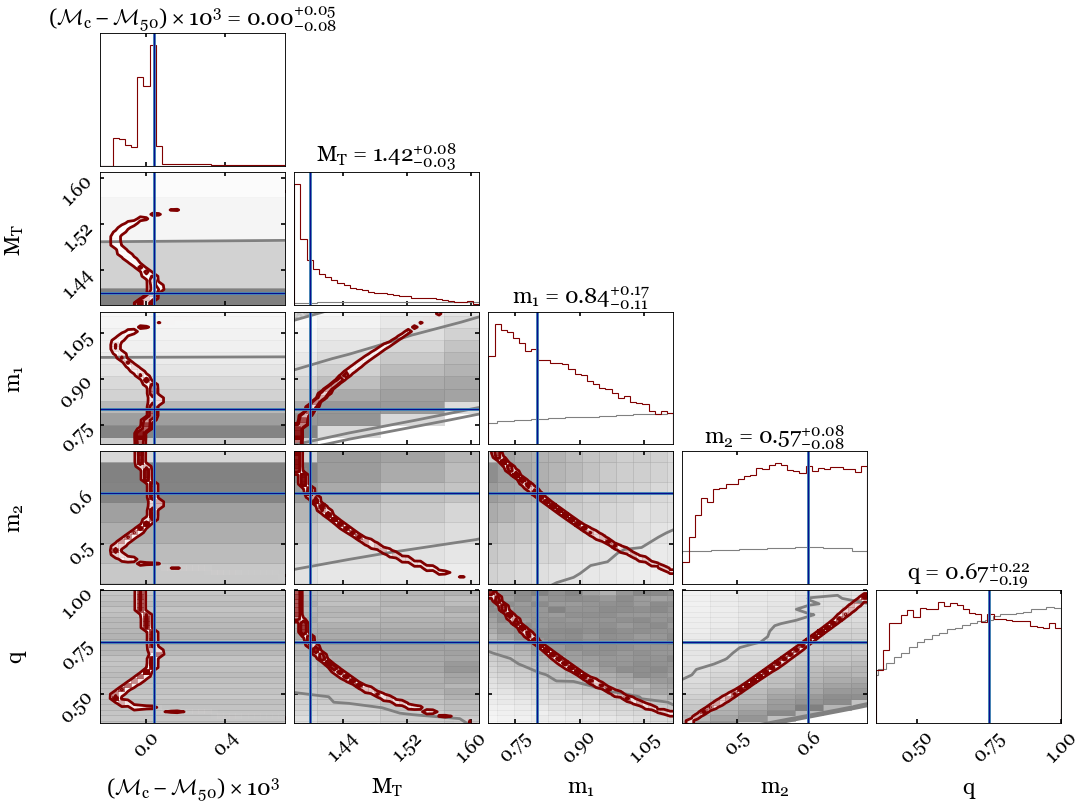}}
    \end{subfigure}
    \caption{(a) Posterior distribution using the power-law fit for $I_{\text{wd}}$ from Kuns et al.~\cite{Kuns_2020} (b) Posterior distribution using the empirical fit for $I_{\text{wd}}$ from Wolz et al.~\cite{Wolz_2020}. The prior distributions (in grey) are plotted over the posteriors (in red). The blue lines represent the injected values for each parameter. All the masses are in units of solar mass ($M_{\odot}$). To represent the errors in $\mathcal{M}_{c}$, we subtracted off the median of our distribution $\mathcal{M}_{50}$ and scaled everything by $10^{3}$. We are able to constrain ($\mathcal{M}_{c}, M_{T}$), but are unable to get good constraints on the corresponding component masses through post-processing, though we still learn information about the component masses compared to our prior distributions. We see that the peaks of the $(\mathcal{M}_{c}, M_{T})$ distributions are offset away from the true values in both models due to the contributions from our prior distributions, and we see a multi-valued distribution in (b) due to the non-monotonic nature of the empirical fit.}
    \label{fig:full posterior}
\end{figure*}

While including tidal effects allows us to detect changes in the frequency evolution beyond the leading order post-Newtonian model, this does not guarantee that the individual masses can be measured. We find that while the component masses are not very well constrained, we are able to get find useful constraints on chirp mass and total mass.
Figure \ref{fig:full posterior} shows the posterior distributions we found for both moment of inertia models, sampling in $(\mathcal{M}_{c}, M_{T})$ and post-processing to then get the component masses $m_{1}, m_{2}$ and corresponding mass ratio $q = m_{2}/m_{1}$. We also overlay the prior distributions (represented by the grey curves) on the posteriors for comparison. We can see that even if our component masses are not well-constrained, the posterior distributions differ from the prior distributions, so information has been gained about the component masses.

\begin{figure}[h!]
    \centering
    {\includegraphics[width=0.45\textwidth]{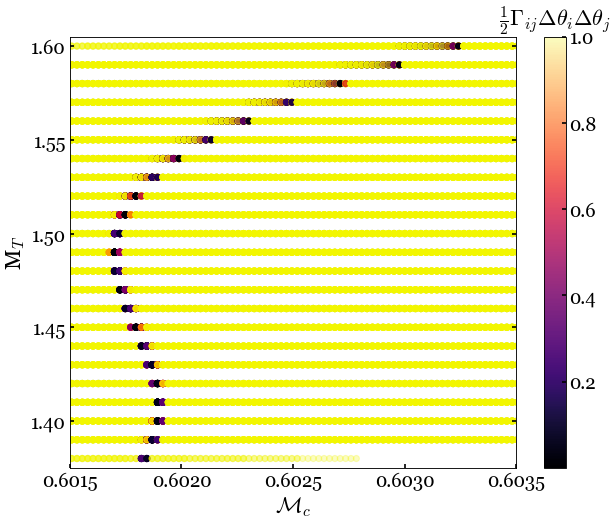}}
    \caption{The log-likelihood evaluated for parameters $(\beta, \delta\gamma)$ for a range of $(\mathcal{M}_{c}, M_{T})$. We implement a threshold and saturate all points that fall above that threshold to isolate mass pairs that give $\beta - \beta_{\text{true}}\simeq0$ and similarly for $\delta\gamma$. We were able to accurately recreate the correlations between $(\mathcal{M}_{c}, M_{T})$ shown in Figure \ref{fig:full posterior}(b).
    }
    \label{fig:color plot}
\end{figure}

There are several features worth mentioning about Figure \ref{fig:full posterior}, as we see significant differences in posterior distributions just by changing the $I_{\text{wd}}$ model. The most notable features are the multi-valued 2D histograms for $\mathcal{M}_{c}$ we see from using the empirical model. We can see based on these 2D histograms that there can be a single value of $\mathcal{M}_{c}$ that produces two different values for the total mass (and the component masses via post-processing). This type of behavior is not captured by a simple Fisher Matrix analysis, as was done in \cite{Wolz_2020}. We think the existence of this multi-valued feature is due to the fact that the empirical fit for $I_{\text{wd}}$ does not monotonically increase. To further probe and validate this multi-valued feature, we created a color plot (Figure \ref{fig:color plot}) calculating the quantity 
\beq
\frac{1}{2}\Gamma_{ij}\Delta\theta_{i}\Delta\theta_{j}
\eeq
for a range of $(\mathcal{M}_{c}, M_{T})$. Here our parameters of interest, $\Delta\theta_{i}$ correspond to to the quantities
\begin{align}
&\beta - \beta_{\text{true}} 
&\delta\gamma - \delta\gamma_{\text{true}}
\end{align}
where $\beta_{\text{true}}, \delta\gamma_{\text{true}}$ are the parameterized frequencies corresponding to our example binary. $\Gamma_{ij}$ is the corresponding Fisher Matrix elements. We use the full expression of the Fisher Matrix here, including the cross-terms, as the correlations between the mass parameters are too large to ignore.

\begin{figure}[h!]
    \centering
    \begin{subfigure}[]
        \centering
        {\includegraphics[width=0.47\textwidth]{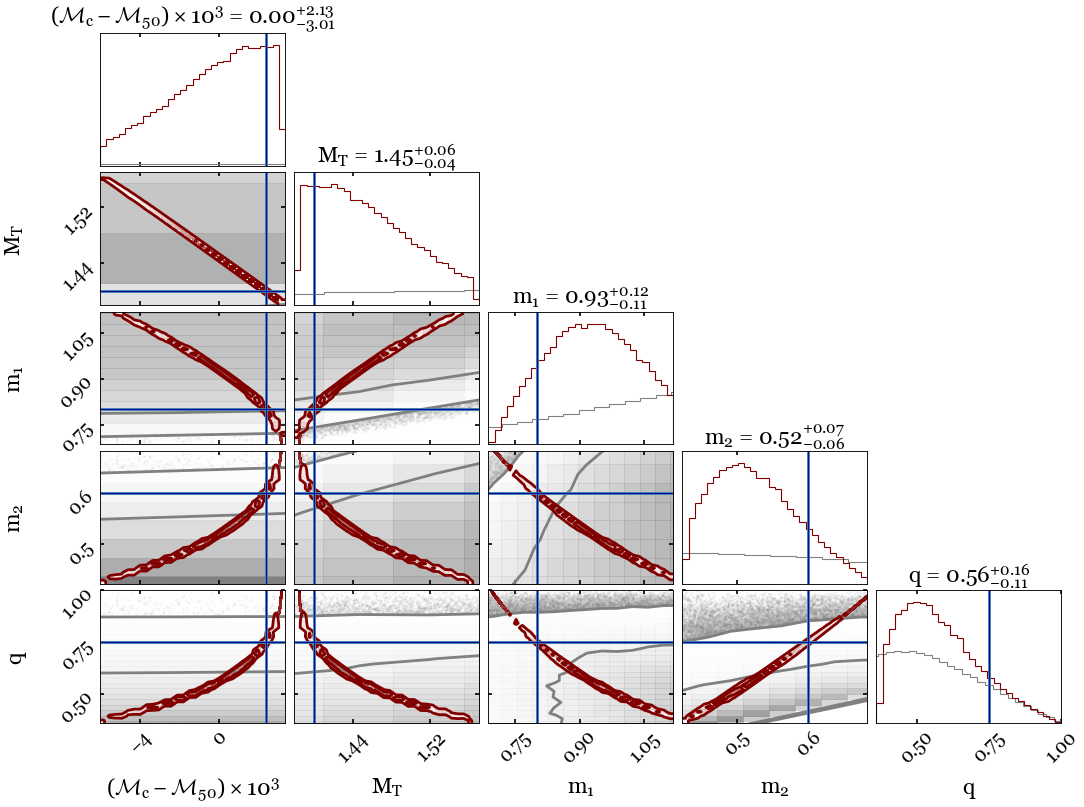}}
    \end{subfigure}
    \begin{subfigure}[]
        \centering
        {\includegraphics[width=0.47\textwidth]{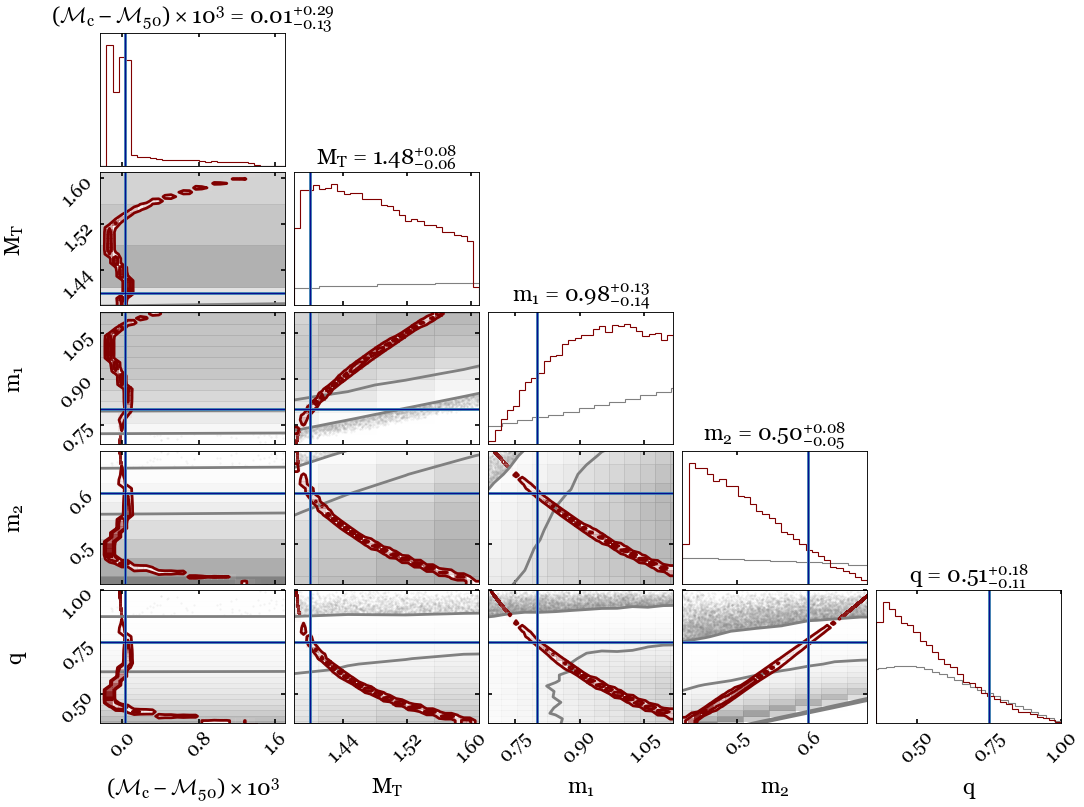}}
    \end{subfigure}
    \caption{Posterior distributions for a frequency evolution using prior distributions that are uniform in $(\mathcal{M}_{c}, M_{T})$. All the masses are in units of solar mass ($M_{\odot}$). The structure changes significantly between this posterior and Figure \ref{fig:full posterior}, where we used priors that were uniform in $(m_{1}, q)$. The prior distributions are overplotted in grey. To represent the errors in $\mathcal{M}_{c}$, we subtracted off the median of our distribution $\mathcal{M}_{50}$ and scaled everything by $10^{3}$. (a) Posterior distribution using the power-law fit for $I_{\text{wd}}$ from Kuns et al.~\cite{Kuns_2020}. (b) Posterior distribution using the empirical fit for $I_{\text{wd}}$ from Wolz et al.~\cite{Wolz_2020}.}
    \label{fig:mcmt prior posterior}
\end{figure}

We specifically searched for mass pairs that produce $(\beta, \delta\gamma)$ values similar to the true parameters, which would  minimize Equation 28. These locations correspond to the $(\mathcal{M}_{c}, M_{T})$ values that populate the 2D histogram we see in Figure \ref{fig:full posterior}. Figure \ref{fig:color plot} shows these locations; we found all of the pairs of $(\mathcal{M}_{c}, M_{T})$ that give a value for Equation 28 beneath a specified threshold of our choosing. We then saturated all of the data points that were above that chosen threshold.

By finely scanning the parameter space, we were able to accurately recreate the correlations between $(\mathcal{M}_{c}, M_{T})$ that we see from our full MCMC analysis. This further confirms that these features we see are valid and accurate.

Other interesting aspects of our posteriors for both models is the non-Gaussianity of the distributions and the offset of the peaks away from the true parameters. Since we are analyzing simulated signals without a noise realization, the likelihood will peak at the true parameter values. Projection effects can shift the maxima away from the true values when looking at 2-dimensional corner plots, but in our case the main cause for the offsets are due to the priors. The expressions below show the gradient of the Jacobian, which give the gradient of the priors on ($\mathcal{M}_{c}, M_{T})$, causing this offset. 
\beq
\frac{\partial J}{\partial \mathcal{M}_{c}} = \frac{10}{9}\frac{(1 + \eta)}{M_{T}\eta^{1/5}(1 - 4\eta)^{3/2}}
\eeq
\beq
\frac{\partial J}{\partial M_{T}} = -\frac{10}{9}\frac{\eta^{2/5}(1 + \eta)}{M_{T}(1 - 4\eta)^{3/2}}
\eeq
From these equations, we can see that the chirp mass will be shifted towards a higher value, while the total mass will be shifted towards a lower value, which is evident in our Figure \ref{fig:full posterior}.

We found that the choice of prior distribution plays a significant role in our analysis. In
Figure~\ref{fig:full posterior} we see that uniform priors in ($m_{1}, q$) (see Figure \ref{fig:prior plot}), shown as light grey lines, translate to highly non-uniform priors in $(\mathcal{M}_{c}, M_{T})$, which shifts the peaks of the posterior distributions for these quantities. If, for comparison, we use a prior distribution that is uniform in $(\mathcal{M}_{c}, M_{T})$, we see drastically different structure in our posteriors. Figure \ref{fig:mcmt prior posterior} shows the posterior distributions we generated using this alternate choice of priors, and it is very clear there are significant differences between these and Figure \ref{fig:full posterior}.

Using the power-law model we have deceptively well-behaved distributions; we are able to get very nice constraints on ($\mathcal{M}_{c}$, $M_{T}$) and even ($m_{1}, m_{2}$). The real difference shows when we use the more accurate fit from Wolz et al.~\cite{Wolz_2020}. It becomes very clear that this choice of priors is not effective in estimating any parameters. We get significantly worse constraints on all the parameters compared to Figure \ref{fig:full posterior}, and the multi-valued nature of $I_{\text{wd}}$ makes constraining $\mathcal{M}_{c}$ impossible. Astrophysically, as stated above, these systems are not described in terms of $(\mathcal{M}_{c}, M_{T})$, making it more accurate to use priors in $(m_{1}, q)$ as well. 

In Figure \ref{fig:full posterior} (and Figure \ref{fig:bias test}), one feature we would like to highlight and justify is the precision to which we can constrain the chirp mass $\mathcal{M}_{c}$. Specifically for high-SNR sources, we can reasonably achieve errors that are as small as $10^{-5}$ for the chirp mass. Ref. \cite{Littenberg_Yunes_2019} shows this in their Figure 4; for an SNR 61 source and an observation time of 1 year, they show constraints on $\mathcal{M}_{c}$ that are on the order of $10^{-4}M_{\odot}$. For an SNR 1000 source such as ours, this would scale to errors on the order of $10^{-5}M_{\odot}$, which is what we see in our posterior distributions.

It is also interesting to note that in both models, we can constrain the lower end of the total mass quite well. While there is extended support shown for total masses greater than the true value ($M_T>1.4\,M_\odot$), we find that there is very little support for $M_T\leq 1.4\,M_\odot$. This shows that GW observations can be a powerful tool in distinguishing super-Chandrasekhar systems from sub-Chandrasekhar ones, hence identifying type Ia progenitors (see Ref. \cite{Maoz_2014} for a review). We also note that the Type Ia progenitor problem is still highly uncertain and a super-Chandrasekhar total mass may not be the necessary condition for Type Ia production~\cite{Shen_2015, Guillochon_2010, Dan_2014, Pakmor_2013, Raskin_2014}.

\subsection{Phenomenological Parameterizations}
\label{sec:phenomparam}
Section \ref{sec:freq_parameterization} described our parameterization of frequencies into unitless quantities. This section describes results we found by performing the analysis with the phenomenological parameters $(\alpha, \beta, \delta\gamma)$. This phenomenological parameterization for DWDs is currently used in the global fit analyses for LISA~\cite{Crowder_2007, Littenberg_2011}. It is also very versatile; the phenomenological model is able to handle many different physical cases (eg. mass transfer \cite{Kremer_2017, Sberna_2021}, tidal effects, and potential third bodies~\cite{Robson_2018}.

However, there are some drawbacks from using the phenomenological model if one wants to map to some physical model by re-sampling the posterior distributions, as the prior distributions can be very different, resulting in a paucity of samples in some regions of the physical parameter space. In our case, post-processing involves converting from $(\alpha, \beta, \delta\gamma)$ to the physical parameterization $(\mathcal{M}_{c}, M_{T}, m_{1}, m_{2}, q)$ via a root-finding algorithm and applying the appropriate Jacobian factor to map from uniform priors in $(\alpha, \beta, \delta\gamma)$ to uniform priors in $(m_{1}, q)$. 

\begin{figure*}
    \centering
    \begin{subfigure}[]
        \centering
        {\includegraphics[width=0.49\textwidth]{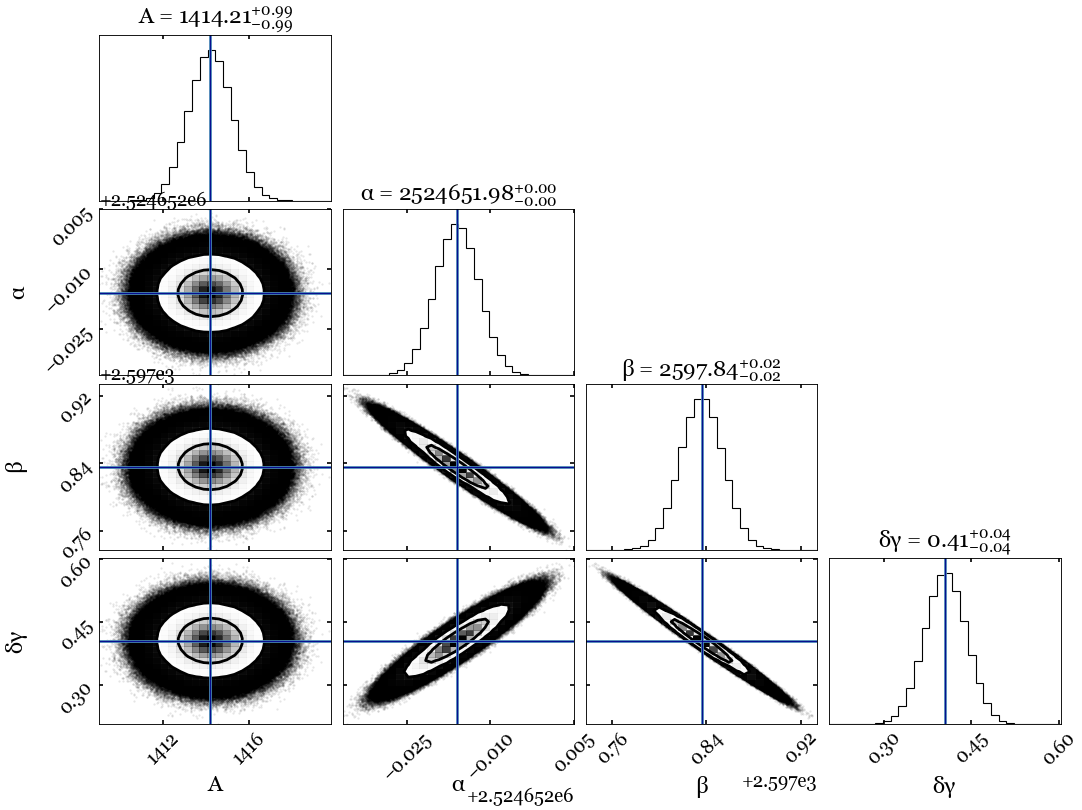}}
    \end{subfigure}
    \begin{subfigure}[]
        \centering
        {\includegraphics[width=0.49\textwidth]{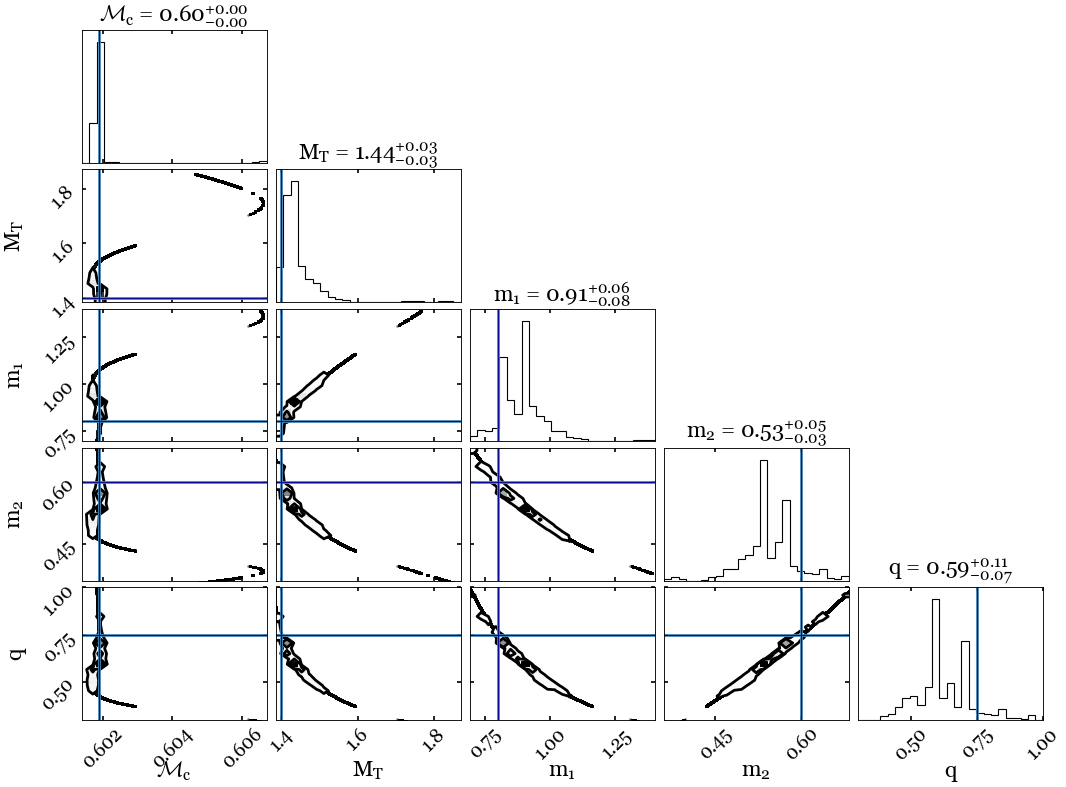}}
    \end{subfigure}
    \centering
    \begin{subfigure}[]
        \centering
        {\includegraphics[width=0.49\textwidth]{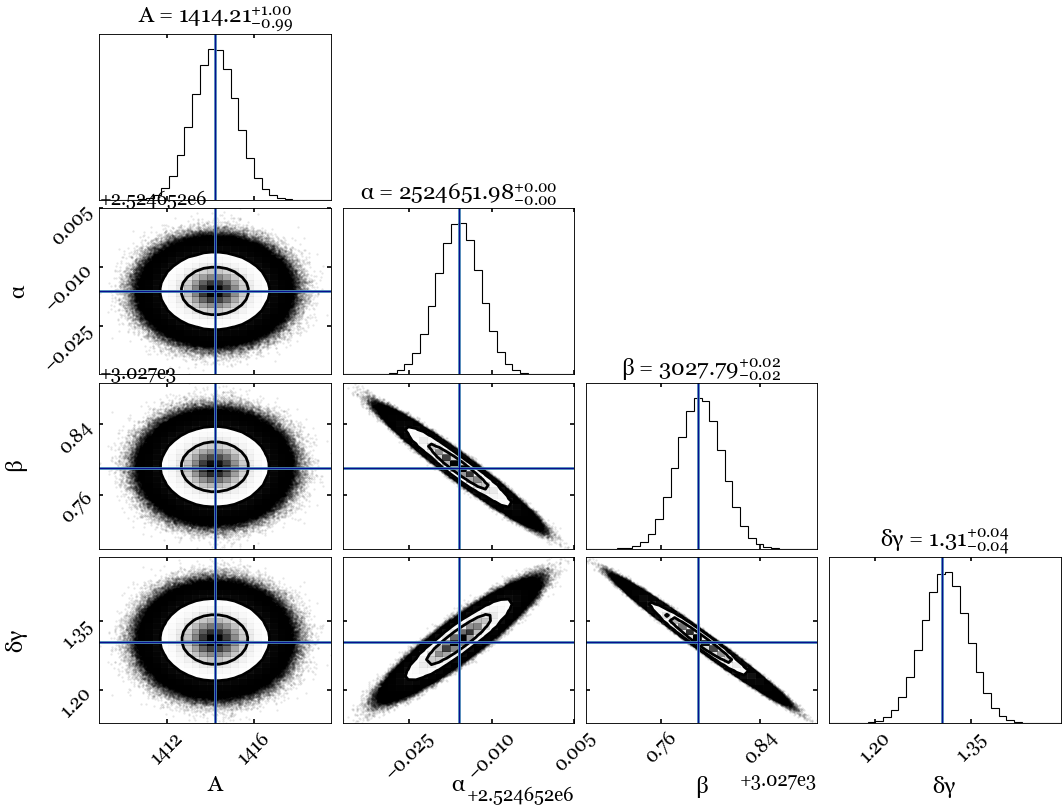}}
    \end{subfigure}
    \begin{subfigure}[]
        \centering
        {\includegraphics[width=0.49\textwidth]{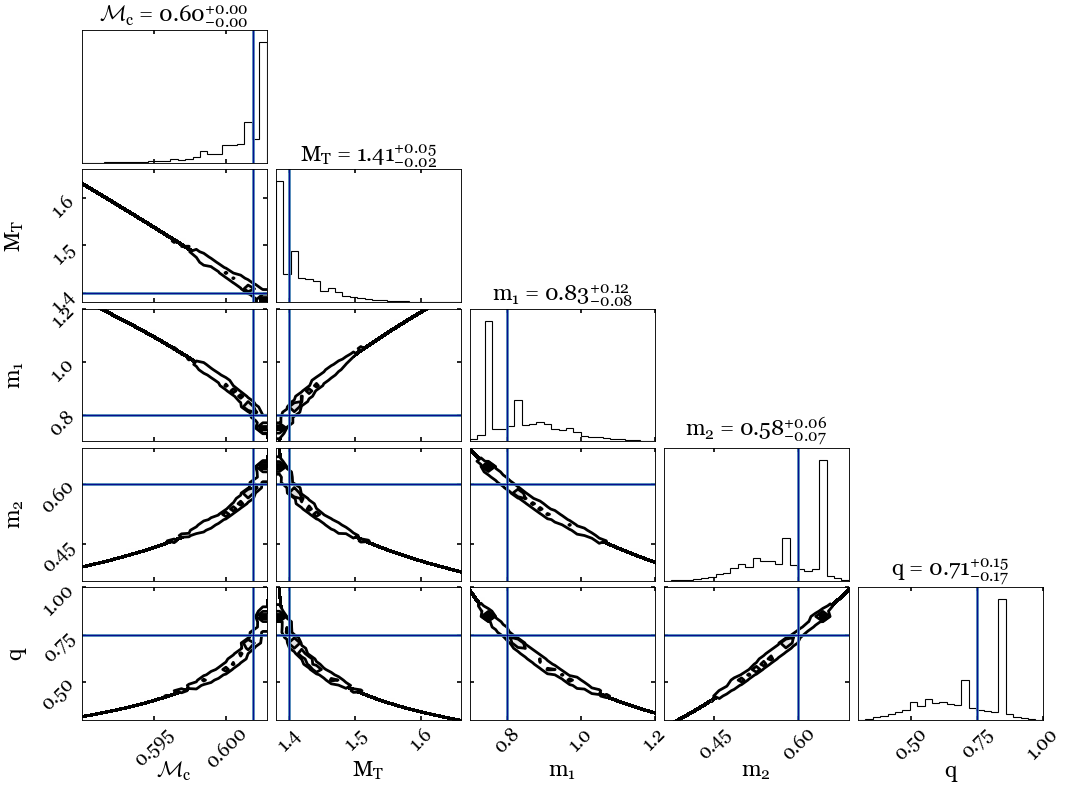}}
    \end{subfigure}
    \caption{Post-processing $(\alpha, \beta, \delta\gamma)$ to the physical parameterization $(\mathcal{M}_{c}, M_{T}, m_{1}, m_{2}, q)$ via a root-finding algorithm and Jacobian mapping from uniform priors in $(\alpha, \beta, \delta\gamma)$ to uniform priors in $(m_{1}, q)$. (a) and (b) are using the empirical model; (c) and (d) are using the power-law model. All the masses (in (b) and (d)) are in units of solar mass ($M_{\odot}$) and the blue lines represent the injected values. Here we see that a significant portion of our frequency data is lost when we post-process, as they do not all map to valid physical parameters.
    } 
    \label{fig:postprocessing}
\end{figure*}

Figure \ref{fig:postprocessing} shows the posteriors we obtained through this post-processing method. We do see that we can capture some of the structure that we see in Figure \ref{fig:full posterior}, but the posterior samples are more sparse. The largest issue to be aware of is the validity of the points being sampled; when sampling in ($\alpha, \beta, \delta\gamma)$, we obtain nice Gaussian distributions centered on our true values. But when we post-process these points to $(\mathcal{M}_{c}, M_{T}, m_{1}, m_{2}, q)$, we lose over 50\% of our original samples as they violate the physical constraints we set on our parameters. 

To be able to obtain posteriors of the same quality as when using a physical parameterization, it requires significantly more computational cost and time. Another weakness that can pose problems with resampling data is that these methods can only transform data that exists; if there are no samples in key regions that you need to explore, post-processing will not fill that gap.

\subsection{Bias Tests}
\label{sec:bias2}

\begin{figure}[h!]
    \centering
    \begin{subfigure}[]
        \centering
        {\includegraphics[width=0.47\textwidth]{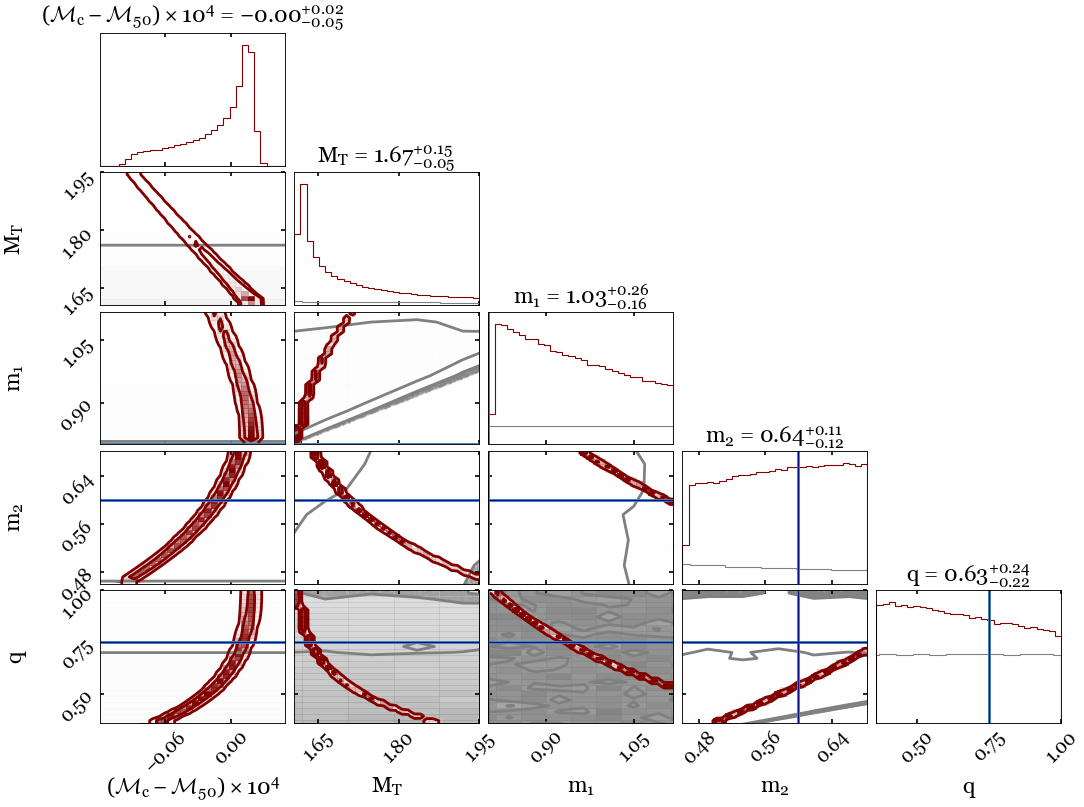}}
    \end{subfigure}
    \begin{subfigure}[]
        \centering
        {\includegraphics[width=0.47\textwidth]{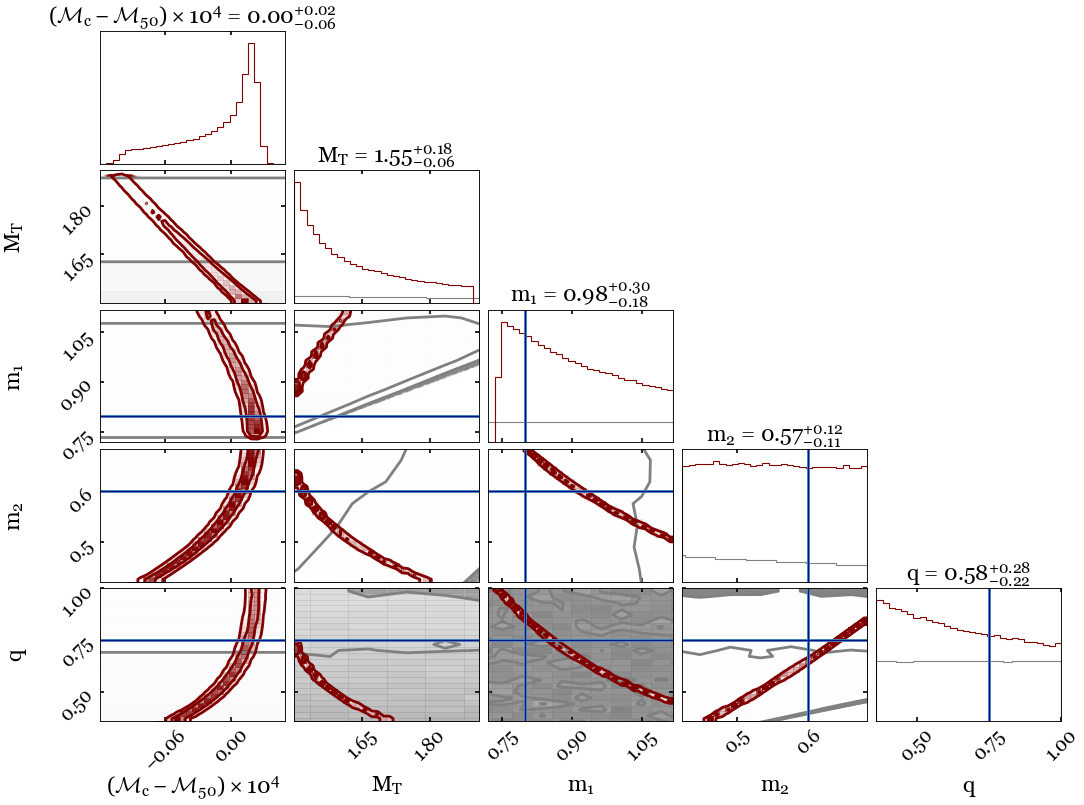}}
    \end{subfigure}
    \caption{Bias tests for a system at a SNR of 1000 and $f = 20$ mHz. The prior distributions (in grey) are plotted over the posteriors (in red). The blue lines represent the injected values for each parameter. All the masses are in units of solar mass ($M_{\odot}$). To represent the errors in $\mathcal{M}_{c}$, we subtracted off the median of our distribution $\mathcal{M}_{50}$ and scaled everything by $10^{4}$. We see for both models the injected values are outside the range of the graph, well beyond the 99\% credible region, marking this as a significant bias towards higher $\mathcal{M}_{c}$ and $M_{T}$.}
    \label{fig:bias test}
\end{figure}

While tidal effects do not allow us to constrain the component masses, they cannot be ignored. Indeed, the opposite is true. Figure \ref{fig:bias test} shows the posterior distributions we get when performing a bias test for our test system at an SNR of 1000 and $f = 20$ mHz. To perform a bias test, we simulate data using the full model (including 1PN and tidal effects) but then we analyze the data with the 0PN point-particle model. 

For our above test system, we see a bias of $\sim10\%$ when using the power-law fit for $I_{\text{wd}}$, and $\sim4\%$ when using the empirical fit. These are extremely statistically significant offsets when compared to how tightly we are able to constrain $\mathcal{M}_{c}$. 

We recognize that our test system parameters of SNR of 1000 and $f = 20$ mHz represents an extreme scenario. As the number density of DWDs in the galaxy scale as $dN/df\propto f^{-11/3}$, we do not expect LISA to detect many systems at this high of an SNR and frequency, though they are not non-existent by any means. Figure 5 in \cite{Cornish_Robson_2017} shows that the five brightest binaries in their analysis all have SNRs above 1000, and Figure 4 shows even more binaries with SNRs above 100. Table 1 of Ref. \cite{Littenberg_Yunes_2019} also quotes, for an observation time of one year, several binaries with SNRs over 100. When extrapolated to an observation time of four years (as SNR grows roughly as $\sqrt{T_{\rm obs}}$), several sources will approach SNRs in the high-hundreds, with one exceeding an SNR of 1000.

Since the bulk of LISA detections will be at much lower SNRs and frequencies, we go on to explore the effect of this bias for systems with SNRs below 50 and frequencies below $f = 10$ mHz. As shown in Figure \ref{fig:beta plot}, we do expect this bias to persist in these low SNR low frequency regimes, which we explore below.  

\begin{figure*}
    \centering
    \begin{subfigure}[]
        \centering
        {\includegraphics[width=0.32\textwidth]{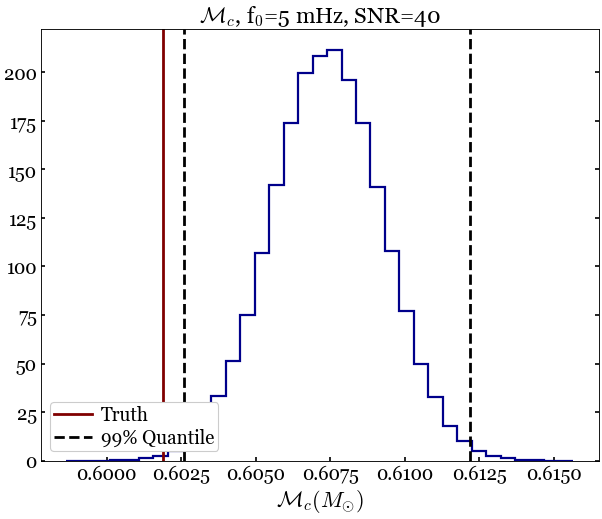}}
    \end{subfigure}
    \begin{subfigure}[]
        \centering
        {\includegraphics[width=0.32\textwidth]{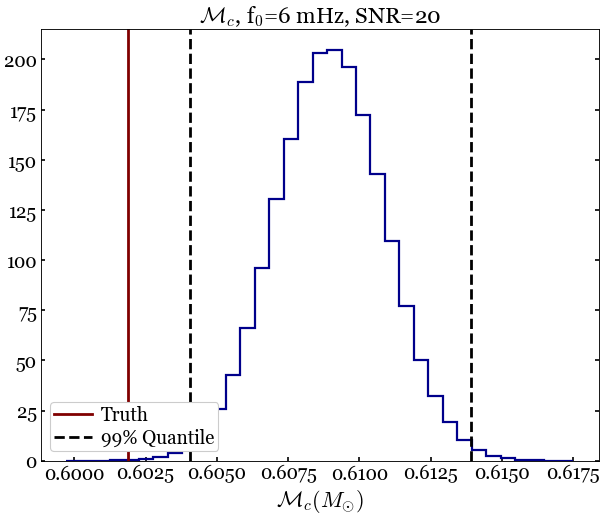}}
    \end{subfigure}
    \begin{subfigure}[]
        \centering
        {\includegraphics[width=0.32\textwidth]{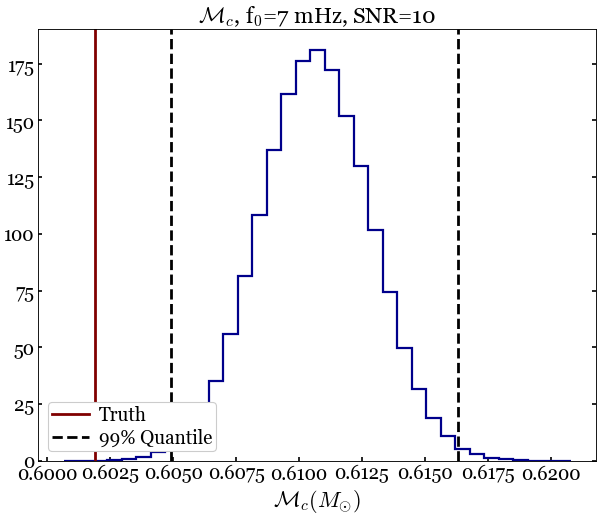}}
    \end{subfigure}
    \caption{For low frequency and SNR systems, we show at what ($f, \text{SNR}$) pairs the bias can no longer be ignored in chirp mass. We define a ``miss'', or bad detection, when the bias shifts the posterior far enough that the true value falls outside the 99\% credible interval.}
    \label{fig:Mc histograms}
\end{figure*}

Though the bias persists through low frequencies, there is a range of frequencies and SNRs where the bias does not shift our posteriors far enough to be statistically significant. Figure \ref{fig:Mc histograms} shows a series of histograms of $\mathcal{M}_{c}$ for a range of low frequencies and SNRs. We explore where the bias pushes the true value outside of the 99\% credible region to determine when this effect is significant. We find that at $f = 5$ mHz, the bias does not shift the distribution outside of the 99\% credible region until an SNR of 40. As we increase the frequency, the SNR we can go up to before this effect happens decreases. By $f = 7$ mHz, this bias is non-negligible at the lowest SNR of 10.

Besides the bias in $\mathcal{M}_c$, we also see a bias in $M_{T}$ towards greater values. In particular, the 10th percentile is greater than the true value by a fraction similar to the fractional bias in $\mathcal{M}_c$. This would cause a sub-Chandrasekhar system to be incorrectly but ``confidently'' identified as a super-Chandrasekhar one.

\hrulefill\\

From these above results, we have two recommendations for the community when dealing with high-frequency sources in LISA: 
\begin{enumerate}
    \item We need to think in terms of the physical parameterization instead of phenomenological, as post-processing and resampling are not sufficient methods for analysis.
    \item Even though these tidal effects are not measurable, they cannot be ignored. We see non-negligible biases from excluding these effects in our analysis, which will significantly change our estimates on $\mathcal{M}_{c}$ and $M_T$. 
\end{enumerate}


\section{Discussion}\label{sec:discussion}

The most significant finding of this study is that the chirp and total mass estimations we get from GW observations can be biased if the tide is not properly accounted for. The bias is approximately $\delta \beta/\beta\sim 1-10\%$ for a DWD in the 10-20 mHz band. If we use such a biased GW sample to constrain parameters in DWD population synthesis studies (such as the common envelope efficiency \cite{Ivanova_2013, Nelemans_2000, Nelemans_2005}), those parameters will further be biased. 

We also demonstrated that the lower bound on $M_T$ is well constrained by LISA. This means a super-Chandrasekhar system (a Type Ia progenitor) can be confidently identified by LISA even if the individual masses are not well constrained. On the other hand, ignoring the tide can lead to a fractional bias in $M_T$ (especially its lower end) similar to the fractional bias in the chirp mass. Such a bias can cause a sub-Chandrasekhar system to be incorrectly identified as a super-Chandrasekhar one, potentially leading to an over estimation of Type Ia rate based on LISA data. 
To reduce biases in both $\mathcal{M}_c$ and $M_T$, improving the accuracy of our waveform models will be crucial. 

An avenue that we have not yet considered is the effect on population studies. From DWD population synthesis models, specific parameters such as common envelope efficiency can potentially be constrained by observations. These parameters will all carry some amount of uncertainty as well. One direction to consider going forward is where the information about these parameters is coming from; will we be able to detect these parameters throughout the bulk of LISA detections? Or will they only be measurable in a few especially well-measure sources? Biases in our modeling can affect how we can constrain and interpret these parameters, which may in turn bias our knowledge of WD formation history.

In this work, we focus on the GW observation of DWDs with LISA alone. There is an exciting possibility that some of the DWDs may also be observed by LSST in the optical band \cite{Korol_2017, Korol_2017_2}. Photometric data from the ongoing Zwicky Transient Facility (ZTF) \cite{Bellm_2019, Graham_2019} and spectroscopic followups from the Extremely Low Mass (ELM) survey \cite{Brown_2010} are two other avenues of potential EM counterparts for these systems. EM observations may directly provide the masses of the WDs through atmosphere modeling \cite{Burdge_2020, Kilic_2021}. If the masses are constrained, GW data then serves as a tool to measure the moment of inertia. Yi et al. 2023 \cite{Yi_2023} shows a complementary analysis using independent measurements of luminosity distance to constrain parameters such as component masses.

Whereas the equilibrium tide (the $\Delta_\Lambda$ term) plays a subdominant role in the GW phase evolution, it can nonetheless leave a strong imprint on the optical signal. As the tide deforms the shape of the WD, it modulates the intensity of the optical light curve, which is known as the ellipsoidal variability. By measuring the modulation depth, one can constrain the radius of the WD. Though to use this information, one also needs to be careful about the modulation produced by the dynamical tide, which can be comparable to the ellipsoidal variability if the WD has a high effective temperature \cite{Yu:21}. 
Alternatively, radius information can be extracted if the binary is eclipsing. Knowing $(m, R, I_{\text{wd}})$ and potentially $\Lambda$ would allow us to accurately constrain a WD's equation of state and test relevant universal relations \cite{Taylor:20}. 

We estimate the tidal effect under the assumption that the WD's spin is always synchronized with the orbit. Multiple studies have found that this condition can be achieved well before a WD enters LISA's band regardless of the details of tidal dissipation~\cite{Fuller_2013, Burkart:13, Yu_2020}. 
Besides spinning up the WD, the tide also generates heat which we ignore in this analysis. Tidal heating can cause additional frequency evolution (e.g., Ref. \cite{Yu_2020})
\beq
    \dot{f}_{\rm heat} \simeq \frac{\Omega_c}{\Omega_{\rm orb}} \dot{f}_{\rm I},
\eeq
where $\dot{f}_{\text{I}}/ \dot f_{\rm pp} =  \Delta_{\rm I} x^2$, $\Omega_{\rm orb}=\pi f_{\rm orb}$ is the orbital angular frequency and $\Omega_c$ is the critical frequency at which the spin-synchronization condition is first achieved. Typically, $2\pi/\Omega_c\simeq \mathcal{O}(10-100)\,{\rm min}$, or $\Omega_c/2\pi \sim 0.1-1\,{\rm Hz}$. Therefore, the change in the frequency evolution rate caused by tidal heating is smaller than spinning up the WD by about one to two orders of magnitude, and their ratio changes as $x^{-3/2}$, so heating is less significant compared to spinning up at higher frequencies. Nonetheless, $\dot{f}_{\rm heat}/\dot{f}_{\rm pp} \propto x^{1/2}$, so it could be potentially important for a very load event at high frequency. 

We also ignore the change in the background structure due to heating. Tidal dissipation is expected to happen near the surface (outer 1\% of the WD radius) where the shear of the tidal wave is the greatest and the wave is most likely to break nonlinearly. Heat generated there does not have enough time to propagate down to the core and significantly alter the moment of inertia during the inspiral~\cite{Yu_2020}. Therefore, $I_{\rm wd}$ should stay as a constant to a reasonable approximation. A more rigorous treatment should couple the orbital evolution with the structural evolution of a WD consistently, similar to recent work \cite{Scherbak:24}, which is left for future investigations.


\section{Conclusion}\label{sec:conclusion}
We studied the effects of tidal interactions in DWDs detectable by LISA in a 4 year observing run. We looked at a large parameter space (e.g., Figs. 4-6, 13) but focused on a $(0.8-0.6)M_{\odot}$ binary at a GW frequency of 20 mHz with an SNR of 1000 for representation.

This paper expands upon and corrects oversights made by Yu et al.~\cite{Yu_2020}, Kuns et al.~\cite{Kuns_2020}, and Wolz et al.~\cite{Wolz_2020}, adjusting the detectability of the moment of inertia, and therefore the component masses, of the system. We find that with these corrections implemented, the moment of inertia can no longer be constrained by better than 1\%, as Yu et al.~\cite{Yu_2020} claimed, nor can the component masses be well constrained at 20 mHz, as Wolz et al.~\cite{Wolz_2020} claimed. We do find, though, that for both models we are able to constrain the lower end of $M_{T}$ very well, showing that using GW observations can be a powerful tool in determining Type Ia progenitors.

We perform a full parallel-tempered Bayesian analysis, including both 1PN and tidal effects. We explore two different models for the moment of inertia. First we consider a scaling relation derived by Kuns et al.~\cite{Kuns_2020}, which is appropriate for lower-mass white dwarfs but loses accuracy past $0.6M_{\odot}$. We then look at an empirical model given by Wolz et al.~\cite{Wolz_2020}, which under-estimates $I_{\text{wd}}$ for low masses, but becomes more accurate at higher masses due to the zero-temperature approximation they used. For the system we are considering in this paper, the empirical model is more accurate, which serves to even further decrease the detectability of $I_{\text{wd}}$ as it has a moment of inertia that is smaller by a factor of $\sim3$ at high masses (Figure \ref{fig:Iwd comp plot}).

We consider two different parameterizations of the frequency evolution -- one version parameterized in terms of physical parameters $(\mathcal{M}_{c}, M_{T})$, and one parameterized in terms of unitless frequency and frequency derivatives $(\alpha, \beta, \delta\gamma)$.

The parameters $\beta$ and $\delta\gamma$ are useful for determining the detectability of the tidal effects and at which frequencies and SNRs the bias starts to appear (Figures \ref{fig:delta plot}, \ref{fig:beta plot}). For our test system, we see that the tidal contribution will not be detectable until we surpass $f =$ 14 mHz using the $I_{\text{wd}}$ from Wolz et al. 

Though it is versatile, we caution against using this phenomenological model for the analysis. When resampling the data from frequencies to masses, we find that there is no guarantee that the corresponding physical parameters are valid; this eliminates a significant section of our dataset when post-processed. To sample enough in frequency to obtain high quality post-processed posteriors requires significantly more time than sampling the physical parameters directly (see Section \ref{sec:phenomparam} and Figures \ref{fig:full posterior}, \ref{fig:mcmt prior posterior}, and \ref{fig:postprocessing}). We also run the risk with resampling of not recovering all the existing features. We cannot resample data in key regions if no samples exist in that region originally.

Each parameterization has its benefits, but overall we find that using a physical model provides greater accuracy in the analysis.

Our most significant result in this paper is shown in Figures \ref{fig:bias test} and \ref{fig:Mc histograms} -- our bias tests. We found that though these tidal features do not lead to detectable component masses, not including them in the model creates significant biases in our detection of the chirp mass and total mass. With our test case, we saw a bias of 10\% higher than the true value when using the scaling model for $I_{\text{wd}}$, and a 4\% shift higher when using the empirical model. 

As our test case is an extreme system in terms of SNR and frequency, we also explored the bias for lower frequency and SNR systems, which will be more numerous. We found that this bias persists through low frequencies and low SNRs, and is still present at $f=5$ mHz and an SNR of 10, which is well below where the tidal contribution would be discernible from high order terms in the frequency evolution (see Figure \ref{fig:Mc histograms}). This low frequency and SNR range will make up a very significant portion of all the LISA galactic binary sources. 

Since $\mathcal{M}_{c}$ is such a well-constrained parameter, these biases are extremely important to take into account. These biases also will have a significant effect on $M_{T}$, potentially leading to false identifications of Type Ia progenitors. So for these LISA sources, we \textit{cannot} neglect the effects due to tides. 


\section{Acknowledgements}
N.J.C. and G.F.  appreciates the support of the NASA LISA Preparatory Science Grant 80NSSC19K0320. H.Y. is supported by NSF grant No. PHY-2308415. We also thank M. Digman for making his LISA parameter estimation code available to check some of our results.


\bibliography{references}


\appendix
\section{Modeling Rotation in White Dwarfs}\label{sec:appendix} 

In Yu et al. 2020 \cite{Yu_2020}, they used a simplified moment of inertia calculation and Fisher Matrix analysis to show that LISA would be able to constrain $I_{\text{wd}}$ to better than 1\% for detached DWDs. In this simplified calculation, they treated the moment of inertia as the parameter of interest. If instead we treat $I_{\text{wd}}=I_{\text{wd}}(m)$, then by measuring the tidal effects we can also measure the component masses of the binary along with the chirp mass. Yu et al.~\cite{Yu_2020} also kept the calculation of $I_{\text{wd}}$ to $0^{th}$ order. This excludes the deformability due to rotation from the expression, which would modify the measurements. 

In Kuns et al. 2020~\cite{Kuns_2020}, they go a step farther and assume a simple fixed mass-moment of inertia relationship (Equation \ref{KunsIwd}), allowing them to obtain uncertainties in inferring the mass ratio for WD binaries with different total masses.

This section is targeted on improving these calculations; we are including the second-order rotational corrections to $I_{\text{wd}}$ while also using a more sophisticated expression than Equation \ref{KunsIwd} to determine if these effects are non-negligible and should be included in our full analysis (see Taylor et al. \cite{Taylor:20} for a complementary analysis of the differential rotation of WDs).

\subsection{Corrections to $I_{\text{wd}}$}
As mentioned above, the rotational speed of a WD will affect its shape. This deformation away from spherical can potentially create a measurable change to the gravitational wave signal emitted.

For the following modifications, we are using the calculations done by Boshkayev et al. 2016 \cite{Boshkayev_2016} and 2017 \cite{Boshkayev_2017}. They apply the Hartle Formalism to slowly-rotating Newtonian configurations, such as WDs, which are to a good approximation non-relativistic. 

The Hartle Formalism~\cite{Hartle_1967}, developed by James Hartle originally for relativistic stars, is a method to simplify the calculations of a rotating star in GR or Newtonian gravity. If a star is rotating slowly enough, the rotation can be considered as a small perturbation onto the equilibrium configuration. This turns the equations of stellar structure into a set of coupled first-order ODEs, simplifying the calculation.

Therefore, the following calculations are done for slowly-rotating WDs that satisfy the following criteria:
\begin{itemize}
    \item A one-parameter equation of state $P=P(\rho)$, where $P, \rho$ are pressure and density.
    \item The non-rotating equilibrium configuration is calculated using the Newtonian equations for stellar structure.
    \item A uniform angular velocity sufficiently slow so that the changes in pressure, energy density, and gravitational field are small. This implies that the angular velocity, $\Omega_{s}$, is
    \beq
    \Omega_{s}^{2} \ll \frac{Gm}{R}
    \eeq
    where $m$ is the unperturbed mass and $R$ is the radius of the star. 
    \item The Newtonian equations are expanded in powers of angular velocity out to second order in $\Omega_{s}$.
\end{itemize}

For this calculation, we considered WDs with a range of masses and radii centered on $m=0.6M_{\odot}$ and $R=10^{9}$ cm. We used a polytropic equation of state with a polytropic index of $n=1.5$, which is a good approximation to the commonly used Chandrasekhar equation of state to leading order \cite{Chandrasekhar_1939}. The equation of state then takes the form $P=K\rho^{5/3}$, where $K$ is a constant.

The $0^{th}$ order stellar structure equations are:
\beq
\frac{dm^{(0)}(r)}{dr} = 4\pi r^{2}\rho (r)
\eeq
\beq
\frac{dP^{(0)}(r)}{dr} = -\rho (r)\frac{Gm^{(0)}(r)}{r^{2}}.
\eeq

To expand in powers of $\Omega_{s}$, we define new coordinates (shown pictorially in Figure 1 of \cite{Boshkayev_2016}): 
\beq
\theta = \Theta \,\,\,\,\,\,\,\,\,\, r = R + \xi (R,\Theta)+O(\Omega^{4}).
\eeq

We then Taylor expand the $0^{th}$ order quantities to second-order and expand $\xi$ in spherical harmonics to simplify the equations. This separates the second-order equations into an $l=0,2$ set of equations.

The spherical part of the deformation corresponds to the $l=0$ equations, which are written as
\beq
\frac{dm^{(2)}(r)}{dr} = 4\pi r^{2}\rho(r) \frac{d\rho(r)}{dp(r)}p_{0}^{*}(r)
\eeq
\beq
\frac{dp_{0}^{*}(r)}{dr} = \frac{2}{3}\Omega_{s} r^{2} - \frac{Gm^{(2)}(r)}{r^{2}}.
\eeq
The equation for total mass becomes $m(r) = m^{(0)}(r)+m^{(2)}(r)$ and likewise for the pressure.

The $l=2$ spherical harmonic corresponds to the quadrupolar deformation of the star. This is originally a second-order ODE, but can be simplified down into two first-order inhomogeneous differential equations:
\beq
\frac{d\chi(r)}{dr} = -\frac{2Gm(r)}{r^{2}}\varphi(r) + \frac{8\pi}{3}\Omega_{s}^{2}r^{3}G\rho(r)
\eeq
\begin{multline}
\frac{d\varphi(r)}{dr} = (\frac{4\pi r^{2}\rho(r)}{m(r)}-\frac{2}{r})\varphi(r) \\- \frac{2\chi(r)}{Gm(r)} + \frac{4\pi\Omega_{s}^{2}r^{4}}{3m(r)}\rho(r).
\end{multline}

Integrating the above equations with their respective boundary conditions will give the particular solutions. To get the homogeneous solutions, we can integrate the homogeneous versions of the equations, where $\Omega_{s} = 0$.

With the same expansion methods that we used for the stellar structure equations, we can obtain the rotational correction terms to the moment of inertia, which are
\beq
I_{\rm wd}^{(0)}(R) = \frac{8\pi}{3}\int^{R}_{0}\rho(r)r^{4}dr
\eeq
and
\beq
I_{\rm wd}^{(2)}(R) = \frac{8\pi}{3}\int^{R}_{0} \left(\left[\frac{1}{5}\xi_{2}(r) - \xi_{0}(r)\right]\frac{d\rho(r)}{dr}\right)r^{4}dr.
\eeq

$\xi_{0}(r)$ and $\xi_{2}(r)$ are given by
\beq
\xi_{0}(r) = \frac{r^{2}}{Gm^{(0)}(r)}p_{0}^{*}(r)
\eeq
and
\beq
\xi_{2}(r) = -\frac{r^{2}}{Gm^{(0)}(r)}\left\{\frac{1}{3}\Omega_{s}^{2}r^{2} + \varphi_{\text{in}}(r)\right\}, 
\eeq
where $\varphi_{\text{in}}$ is obtained through matching solutions for $\varphi$ on the boundary $r=R$.

\begin{figure}[h!]
    {\includegraphics[width=0.45\textwidth]{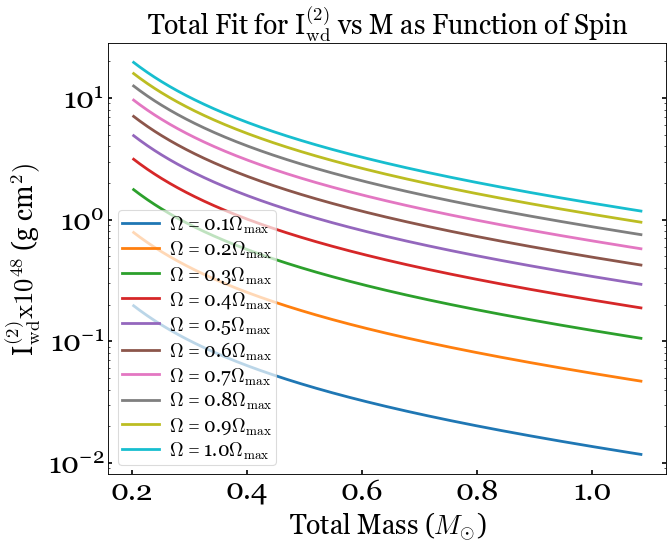}}
    \caption{The total fit for the moment of inertia correction for changing mass and angular frequency. It can be fit using a product of polynomials shown in (A13).}
    \label{fig:I2 fit}
\end{figure}
\begin{figure}[h]
    {\includegraphics[width=0.45\textwidth]{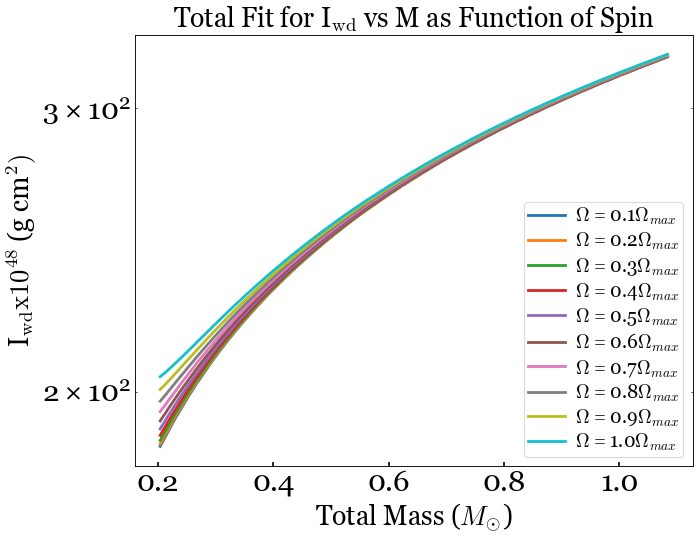}}
    \caption{The total fit for the total moment of inertia. The rotation will only play a significant role in modifying $I_{\text{wd}}$ for low mass stars.}
    \label{fig:Itot fit}
\end{figure}

To look at how the moment of inertia changes with spin and mass, we found that the second order correction to the moment of inertia $I^{(2)}_{\text{wd}}(\Omega_{s}, m)$ increased by a scaling factor $\alpha$ multiplied by $\Omega_{\rm max}^{2}$ for each change in angular frequency $\Omega_{s}$. Here, $\Omega_{\rm max}$ is the maximum spin that we allow the star to reach.

By fitting the moment of inertia with a polynomial fitting function, and similarly for the scaling factors, we were able to obtain a total fit for $I_{\text{wd}}(\Omega_{s}, m)$ from the product of the two individual fits:

\begin{multline}
F(\Omega_{s}, m) = \frac{I^{(2)}(\Omega_{s}, m)}{\alpha\Omega_{\rm max}^{2}(\Omega_{s})} = \frac{a_{0} + a_{1}M^{-1} + a_{2}M^{-2}}{b_{0}\Omega^{-2}}. 
\end{multline}

The fitting parameters are $a_{0} = -0.00563, a_{1} = 0.01399 M_{\odot}, a_{2} = 0.00584 M_{\odot}^{2}, b_{0} = 1.02838 \unit{rad^{2}/s^{2}}$. Figure \ref{fig:I2 fit} shows this fit for a range of angular frequencies from $0.1\Omega_{\rm max}$ to $\Omega_{\rm max}$, where $\Omega_{\rm max} =0.0314 \unit{rad/s}$. This corresponds to a rotational frequency of $f = 10\unit{mHz}$, or a GW frequency of $f = 20 \unit{mHz}$. We select this maximum spin to correspond with the test system used in this paper, which emits GWs at $f = 20 \unit{mHz}$.

To see how spin contributes to the overall moment of inertia, we added in the non-rotating component, $I^{(0)}_{\text{wd}}$, shown in Figure \ref{fig:Itot fit}. With the $0^{th}$ order term added in, it becomes clear that the rotation only will have a significant effect on low mass white dwarfs as the higher the mass, the harder it is to perturb the star.

\end{document}